\documentclass[
 reprint,
superscriptaddress,
 amsmath,amssymb,
 aps,
floatfix
]{revtex4-2}

\usepackage{graphicx}
\usepackage{dcolumn}
\usepackage{bm}
\usepackage{hyperref}

\usepackage[most]{tcolorbox}
\usepackage{mathtools}
\usepackage{amsmath}

\newcommand{\mS}{\mathcal{S}}
\newcommand{\mL}{\mathcal{L}}

\newcommand{\mO}{\mathcal{O}}

\newcommand{\mn}{{\mu\nu}}

\newcommand{\rs}{{\rho\sigma}}
\newcommand{\ab}{{\alpha\beta}}
\newcommand{\ba}{{\beta\alpha}}
\newcommand{\gd}{{\gamma\delta}}
\newcommand{\dg}{{\delta\gamma}}

\newcommand{\Poincare}{Poincar\'e }
\DeclareMathOperator{\Tr}{Tr}

\newcommand{\IS}[3]{
\underbracket[0.5pt][2pt]{#1}_{([#2][#3])}
}

\begin{document}


\title{On the role of Area Metrics in AdS/CFT}

\author{Aranya Bhattacharya}
 \email{aranya.bhattacharya@bristol.ac.uk}
\affiliation{%
Institute of Theoretical Physics, Jagiellonian University, \\
{\L}ojasiewicza 11, 
30-348 Krak{\'o}w, Poland
}
\affiliation{School of Mathematics, University of Bristol, Fry Building \\
Woodland Road, Bristol BS8 1UG, UK}
\author{Lavish Chawla}%
\email{lavish.lavish@doctoral.uj.edu.pl}
\affiliation{%
Institute of Theoretical Physics, Jagiellonian University, \\
{\L}ojasiewicza 11, 
30-348 Krak{\'o}w, Poland
}
\affiliation{Theoretisch-Physikalisches Institut, Friedrich-Schiller-Universit\"at Jena, \\ Max-Wien-Platz 1, D-07743 Jena, Germany}
\author{Mario Flory}
 \email{mflory@th.if.uj.edu.pl}
\author{Mateusz Kulig}
\email{mateusz.1.kulig@student.uj.edu.pl}
 
\affiliation{%
Institute of Theoretical Physics, Jagiellonian University, \\
{\L}ojasiewicza 11, 
30-348 Krak{\'o}w, Poland
}%


\begin{abstract}
We explore the potential role of area metrics, a generalised notion of geometry, in the AdS/CFT correspondence. Guided by the Ryu-Takayanagi formula and the first law of entanglement, we derive both a holographic dictionary as well as a bulk equation of motion for area metric fluctuations linearised around a four dimensional AdS background. Furthermore, we explore potential bulk Lagrangians that could give rise to the derived equation. It is shown that writing area metric fluctuations in terms of Lanczos like potentials gives rise to a particularly elegant bulk model. 
\\

\noindent
A 15-minute video abstract can be found at \href{https://www.youtube.com/watch?v=SWBCICeFhbw}{this link}
\end{abstract}

\maketitle


\section{Introduction}
\label{sec::Intro}

There are many reasons to think that the concept of \textit{area} might be more fundamental in quantum gravity than the concept of \textit{length}. For instance, it is the area of a worldsheet that defines the Nambu-Goto action of bosonic classical string theory \cite{Goto:1971ce}, area variables play an important role in Loop Quantum Gravity (LQG) \cite{Rovelli:1994ge,Ashtekar:1996eg,Dittrich:2008ar,Perez:2012wv,Dittrich:2008va,Asante:2020qpa,Asante:2020iwm}, and black hole entropy \cite{Bekenstein:1972tm,Bekenstein:1973ur} as well as holographic entanglement entropy \cite{Ryu:2006bv,Ryu:2006ef} are given by areas measured in squares of the Planck length. 
In fact, one can even go so far as \textit{deriving} Einstein's equations of motion as an emergent phenomenon by equating the notions of area and entropy as demonstrated in \cite{Jacobson:1995ab}. 
In the context of the AdS/CFT correspondence, it was likewise shown in \cite{Blanco:2013joa,Lashkari:2013koa,Faulkner:2013ica} how to derive the linearised Einstein's equations around the AdS background from the Ryu-Takayanagi formula \cite{Ryu:2006bv,Ryu:2006ef} and the first law of entanglement in the boundary theory. Depending on one's perspective, this result might be seen as a consistency check on the established AdS/CFT correspondence, or as an additional example of the emergence of gravity in the tradition of \cite{Jacobson:1995ab}. See \cite{deHaro:2015uta,DeHaro:2015aht} for a discussion on the difference between the ideas of \textit{duality} and \textit{emergence}.

Interestingly, there is a generalised type of geometry called \textit{area metrics} that allows to define the area
\begin{align}
A[\Sigma]=\int_\Sigma dA 
\end{align}
of a 2-surface $\Sigma$ via an area element
\begin{align}
dA^2=\frac{1}{4}G_{\mu\nu\rho\sigma}\left(dx^\mu\wedge dx^\nu\right)\otimes \left(dx^\rho\wedge dx^\sigma\right),
\label{area_element}
\end{align}
just like an ordinary metric defines the length $L[\gamma]=\int_\gamma ds $ of a curve $\gamma$ via the line element $ds^2=g_{\mu\nu}dx^\mu dx^\nu$ \cite{Reisenberger:1998fk,Schuller:2005yt}. 
Equation \eqref{area_element} requires the area metric $G_{\mu\nu\rho\sigma}$ to have the index symmetries \cite{Schuller:2005yt}
\begin{align}
G_{\ab\gd}&=-G_{\ba\gd}=-G_{\ab\dg}=G_{\gd\ab}.
\label{symm1}
\end{align}
In addition, some (e.g. \cite{Schuller:2005yt}, but not all, e.g. \cite{Schuller:2005ru}) papers require a cyclicity condition
\begin{align}
\epsilon^{\ab\gd}G_{\alpha\beta\gd}=0
\label{cyclicity}
\end{align}
which we will also assume from now on. Essentially, because of \eqref{symm1} and \eqref{cyclicity}, the area metric looks like a Riemann-tensor.

In $d\geq4$ dimensions, area metrics describe more general notions of geometry than ordinary metrics, as any length metric induces an area metric via the formula 
\begin{align}
G^{(g)}_{\mu\nu\alpha\beta}=g_{\mu\alpha}g_{\nu\beta}-g_{\nu\alpha}g_{\mu\beta},
\label{InducedAreametric}
\end{align}
but not every area metric is induced by a length metric in this way \cite{Schuller:2005yt}. See table \ref{tab:dofs} for a counting of free components of both metrics and area metrics based on the number of dimensions $d$.

There is a considerable body of literature on area metrics \cite{Schuller:2005yt,Punzi:2006nx,Punzi:2006hy,Punzi:2007di,Schuller:2007ix,Punzi:2008dv,Punzi:2009yq,Schuller:2009hn,Dahl:2011hb,Ho:2015cza}, and they have made appearances in various approaches to (quantum) gravity including (but not limited to, see \cite{Kuipers:2024gfp,Kuipers:2025gux,Bianconi:2025rnd}) string theory \cite{Schuller:2005ru,Borissova:2024cpx}, LQG \cite{Dittrich:2022yoo,Dittrich:2023ava,Borissova:2023yxs,Davidson:2024nlh} and asymptotic safety \cite{Borissova:2025frj}, but to our knowledge this is the first paper that systematically studies their potential role within the AdS/CFT correspondence. 
The motivation for this is that in AdS/CFT, only special states of the CFT actually have a valid description in terms of a classical Einsteinian bulk geometry. 
By introducing a more general notion of geometry for the bulk gravity,  we hope that a larger subset of the Hilbert space of the putative dual theory can be studied holographically. 
In particular, it has been suggested \cite{Schuller:2005yt,Ho:2015cza} that area metrics might serve to encode (quantum) superpositions or ensemble averages of ordinary metric geometries. 

Concretely, we will follow the techniques of \cite{Blanco:2013joa,Lashkari:2013koa,Faulkner:2013ica} to show in a first principles manner how area metrics can fit into the AdS/CFT correspondence by deriving their holographic dictionary entry and their bulk equations of motion. To achieve this,  we will fix $d=4$ throughout and study linearised area metric like fluctuations around a fixed AdS$_4$ background $g_\mn$ \footnote{In what follows, as a convention, we take $a,b$ indices as boundary spacetime indices, and $\mu,\nu$ Greek indices to denote the bulk spacetime indices. Finally, for some expressions when we only need the boundary spatial indices, we denote them by $\vec{x}$ or $x^I$.}, 
\begin{align}
G_{\mn\ab}&\equiv g_{\mu\alpha}g_{\nu\beta}-g_{\nu\alpha}g_{\mu\beta} + a_{\mn\ab},\ \ (a\ll1),
\label{AMfluctuation}
\\
g_{\mn}dx^\mu dx^\nu&\equiv \frac{L^2}{z^2}\left(dz^2+\eta_{mn}dx^m dx^n\right).
\end{align}
These area metric fluctuations can be split as \cite{Borissova:2023yxs}
\begin{align}
 &  a_{\mn\ab}  =2 h_{\alpha[\mu}g_{\nu]\beta}-2 h_{\beta[\mu}g_{\nu]\alpha} +w_{\mn\ab}.
 \label{fluctuationdecomp}
\end{align}
where the first two terms on the right hand side correspond to a metric-like fluctuation $g_\mn\rightarrow g_\mn+h_\mn$, and the last term $w_{\mn\ab}$ is a purely area metric-like fluctuation, satisfying 
\begin{align}
w^{\mu}_{\ \nu\mu\delta}=0,\ \epsilon^{\ab\gd}w_{\ab\gd}=0.
\label{TraceW}
\end{align}
Simply speaking, if $a_{\mn\ab}$ looks like a Riemann tensor, then $w_{\mn\ab}$ looks like a Weyl tensor. Out of the 20 components of the area metric fluctuation that have to be specified (see table \ref{tab:dofs}), 10 are encoded in $h_{\ab}$ and 10 in $w_{\mn\ab}$. 
The general area metric fluctuation $a_{\mn\ab}$ can actually be split in a finer way by separating trace and traceless part of $h_{\mn}$ as well as self-dual and anti-self-dual parts of $w_{\mn\ab}$ \cite{Borissova:2023yxs}, but this will not be necessary in this paper.

\begin{table}[tb]
        \caption{We show how many free components can be specified as a function of the dimension $d$ for a metric or area metric (where \eqref{symm1} and \eqref{cyclicity} hold) \cite{Ho:2015cza}. This does not take into account gauge redundancy or the number of physical degrees of freedom. 
        In $d=2$, an area metric encodes the same information as the volume element, and so has only one free component. In $d=3$, metric and area metric geometry are equivalent and can be transformed into each other \cite{Ho:2015cza}. In $d\geq4$, area metrics represent a more general type of geometry than ordinary metrics. }
    \label{tab:dofs}
    \begin{ruledtabular}
    \begin{tabular}{c|ccccc|c|c}
       $d$ &2 & 3 & 4 & 5 & 6  & 10 & 26 \\
       \hline
      $g_\mn$&3 & 6  & 10 & 15 & 21  &55  & 351\\
      $G_{\ab\gd}$ &1  & 6 & 20 & 50 & 105  & 825 & 38025 \\     
    \end{tabular}
    \end{ruledtabular}
\end{table}

\section{Existence of dual EMT}
\label{sec::ExistenceEMT}

Without a doubt, area metrics are a considerable departure from the familiar (length-) metric based Riemannian geometry. If such a generalised notion of geometry is introduced in the bulk, does that mean that the (putative) dual theory is likewise only defined on a generalised area metric background? 

The answer to this is no, for the following reason: Suppose we look at a near-boundary cutoff $z=\epsilon$.
Here, $\epsilon$ doesn't even need to be small. 
This defines a co-dimension one surface in the bulk, on which the bulk area metric $G_{\mn\rs}$ defines an \textit{induced area metric} $G_{mnrs}$. 
But a co-dimension one surface in the bulk is three-dimensional, and in three dimensions area metrics and ordinary metrics are interchangeable, see table \ref{tab:dofs} and \cite{Ho:2015cza}. One can translate from one to the other using \eqref{InducedAreametric} and 
\begin{align}
g^{ab}
=\mp\frac{1}{8}
\epsilon_{opq}\epsilon_{rst}
G^{aobr}G^{pqst}. 
\end{align}
Hence the boundary theory can be equivalently written as either defined on a metric background or an area metric background in terms of an action 
\begin{align}
\mS=\mS[g_{mn},\phi,\partial_a\phi,...]=\mS[G_{mnrs},\phi,\partial_a\phi,...]
\end{align}
where $\phi$ are the fields of the theory. This allows us to define not only an \textit{energy momentum tensor} (EMT) 
\begin{align}
T_{ab}=\frac{-2}{\sqrt{-g}}\frac{\delta \mS}{\delta g^{ab}},
\label{Tab}
\end{align}
for the dual field theory, but also a new tensor
\begin{align}
T_{abcd}\equiv\frac{-2}{\sqrt{-g}}\frac{\delta \mS}{\delta G^{abcd}}=T_{mn}\frac{\delta g^{mn}}{\delta G^{abcd} }.
\label{Tabcd}
\end{align}
which we term the \textit{surface energy tensor}. 
It will always be possible to translate $T_{abcd}$ to $T_{mn}$ and vice versa, i.e.~these two tensors encode the same physical information in different ways. See the supplementary material for detailed expressions.

\section{Holographic Dictionary}
\label{sec::Dictionary}

In the context of the AdS/CFT correspondence, the Ryu-Takayanagi (RT) formula \cite{Ryu:2006bv, Ryu:2006ef} relates the entanglement entropy of a subregion $A$ of a pure state in the boundary CFT to the area of a minimal surface $\Sigma$ in the dual bulk spacetime, anchored at the boundary $\partial A$ of the original subregion in the CFT:
    $S_A=\frac{\text{Area}(\Sigma)}{4G_N}.$
In the field theory, this observable is computed by the von Neumann entropy $S_A=-\Tr(\rho_A \log\rho_A)$ of the reduced density matrix $\rho_A$ obtained by tracing out the complementary subregion from the full density matrix $\rho$.

The central relation underlying our work is the \textit{first law of entanglement} \cite{Bhattacharya:2012mi, Blanco:2013joa}.
This is an equality 
\begin{equation}
    \delta S_A= \delta\langle H_A\rangle
    \label{first_law}
\end{equation}
between the leading order changes in the entanglement entropy ($\delta S_A$) and the expectation value of the modular Hamiltonian ($\delta \langle H_A \rangle$) when perturbing the overall state as $\rho\rightarrow \rho+\delta \rho$. 
Via the RT formula, this equation relates changes of the bulk geometry to changes of the modular Hamiltonian in the boundary theory. 

The modular Hamiltonian $H_A=-\log(\rho_A)$, which is in general tricky to compute, is known for a ball-shaped boundary spatial subregion centred at $\vec{x}_0$ with radius $R$ on a constant time-slice $t=t_0$ \cite{Bhattacharya:2012mi,Lashkari:2013koa}:
\begin{equation}
\label{defdeltaH}
    \delta \langle H_A\rangle =2\pi \int_{A(R, \vec{x}_0)} d^{2}x \frac{R^2-|\vec{x}-\vec{x}_0|^2}{2R} \delta \langle T_{tt}(t_0, \vec{x})\rangle,
\end{equation}
where $T_{tt}(t_0, \vec{x})$ is the $\{t, t\}-$component of the boundary EMT. In the limit $R\ll 1$, $\langle T_{tt}(t_0, \vec{x})\rangle$ is approximately constant throughout the infinitesimal ball resulting in 
\begin{align}
    &\delta \langle H_A\rangle\big|_{R\ll 1} \approx \frac{\pi^2 R^3 }{2} \delta \langle T_{tt}(t_0, \vec{x}_0) \rangle.
\end{align}
Using \eqref{first_law}, we can then write the expectation value of the EMT for an infinitesimal ball in terms of the variation of entanglement entropy as \cite{VanRaamsdonk:2016exw}
\begin{equation}
\label{SEtensorEE}
    \delta \langle T_{tt}(t_0, \vec{x}_0)\rangle=\frac{2}{\pi^2} \lim_{R\rightarrow 0} \left(\frac{\delta S_{A(R,\vec{x}_0)}}{R^3}\right).
\end{equation}
In our setup, we assume that the RT formula remains valid if the bulk geometry is given by an area metric \eqref{AMfluctuation}, and hence $\delta S_A$ gets contribution from both $h_{\mn}$ and $w_{\mu\rho\nu\lambda}$. Using this in the infinitesimal ball limit \eqref{SEtensorEE}, we derive a holographic dictionary relation between the boundary stress tensor and the perturbations of the bulk geometry. 
By applying a boost along the boundary directions, we obtain a more general dictionary involving the components $\delta \langle T_{ab}\rangle$.  

\paragraph{Known result; $w_{\mu\rho\nu\lambda}$ turned off:}
In the case of usual length metric perturbations $h_{\mn}$, the above relation in the infinitesimal ball limit is well known:
\begin{align}
    \delta \langle T_{ab}(t_0, \vec{x}_0)\rangle=\frac{3}{16\pi G_N} H^{(0)}_{ab}(t_0, \vec{x}_0),
    \label{hDictionary}
\end{align}
where we assume the Fefferman-Graham gauge $h_{\mu z}=0$ throughout the paper and $h_{\mn}=z\sum_{n=0}^{\infty} z^n H_{\mn}^{(n)}(t,\vec{x})$. 
This expansion tells us that the leading order metric perturbation in powers of  $z$ has to be $\sim z^{1}$ for the boundary EMT in the infinitesimal ball limit to be non-vanishing and non-divergent. 
It is worth noting that the tracelessness and conservation of the boundary stress tensor implies the following conditions on $H_{ab}^{(0)}$ \cite{Lashkari:2013koa, Faulkner:2013ica}:
\begin{equation}
H^{(0)}{}_{a}^{a}=0,\,\, \partial^{a}H^{(0)}_{ab}=0.
\end{equation}

\paragraph{New result; $h_{\mn}$ turned off:}
In this case, we turn off the length metric $h_{\mn}$ and keep only the purely area metric fluctuations $w_{\mu\rho\nu\lambda}$. Firstly, the leading order variation in the area of the minimal surface $\Sigma$ takes the form (see appendix \ref{sec::holodicti} for details)
\begin{equation}\label{AMeom}
    \delta A= \int_{\Sigma} d^2 x \frac{ w_{\mu\rho\nu\lambda}\vartheta^{I\,J}\vartheta^{K\,L} \partial_{I} X^\mu \partial_{J} X^\nu \partial_{K} X^\rho \partial_{L} X^\lambda}{8\sqrt{\hat{g}}},
\end{equation}
where $\hat{g}$ is the determinant of the unperturbed induced metric on $\Sigma$ and $\vartheta^{IJ}$ is the Levi-Civita symbol in two dimensions, with $I\in\{x,y\}$. The extremal surface for the leading order variation remains the same as the one for the empty AdS$_4$ \cite{Ryu:2006ef, Bhattacharya:2012mi, Bhattacharya:2019zkb}:
\begin{equation}
    z(x,y)=\sqrt{R^2-|\vec{x}-\vec{x}_0|^2}.
\end{equation}
Evaluating \eqref{AMeom} in the infinitesimal ball limit, we get the following identity for the boundary stress tensor:
\begin{equation}
    \delta \langle T_{ab}(t_0, \vec{x}_0)\rangle=\frac{1}{16\pi L^2 G_N} \eta^{ij}W^{(0)}_{aibj}(t_0, \vec{x}_0),
    \label{wDictionary}
\end{equation}
and also the fact that the $z$ power series expansion of the $w$ fluctuations has to be of the form 
\begin{equation}
    w_{\mu\rho\nu\lambda}(z, t, \vec{x})\equiv\frac{1}{z}W_{\mu\rho\nu\lambda}(z, t, \vec{x})=\frac{1}{z}\sum_{n=0}^{\infty} z^n W_{\mu\rho\nu\lambda}^{(n)}(t, \vec{x}),
    \label{w_as_sum}
\end{equation}
for the stress tensor to have a non-vanishing and non-divergent contribution in the infinitesimal ball limit.

These are the two main results that we obtain as our entry to the holographic dictionary for purely area metric fluctuations. One reassuring thing is to note that the tracelessness \eqref{TraceW} already ensures that the boundary stress tensor sourced by the $w$ fluctuation is traceless. Subsequently, the conservation of the stress tensor implies
\begin{equation}
    \partial^a (\eta^{ij} W^{(0)}_{aibj})=0.
    \label{conserv_w_0}
\end{equation}

\paragraph{Full area fluctuation:}
Finally, when we have the full fluctuation $a_{\mu\rho\nu\lambda}$ turned on, the dictionary takes the form
\begin{align}
    &\delta \langle T_{ab}\rangle=\frac{1}{16\pi G_N} \left(3 H^{(0)}_{ab}+\frac{1}{L^2}\eta^{ij}W^{(0)}_{aibj}\right).
    \label{hwDictionary}
\end{align}
Tracelessness and conservation of the stress tensor then imply
\begin{equation}
    H^{(0)}{}^a_a=0,\, \partial^a\left(3H^{(0)}_{ab}+\frac{1}{L^2}\eta^{ij}W^{(0)}_{aibj}\right)=0.
\end{equation}
This analysis confirms that we can understand the boundary stress tensor  $\langle T_{ab}\rangle$ to source the full area metric fluctuations $a_{\mu\rho\nu\lambda}$ in the AdS$_4$ bulk, not just the $h_\mn$ part. 

\section{Equations of Motion}
\label{sec::EOMs}

It was shown in \cite{Blanco:2013joa,Lashkari:2013koa,Faulkner:2013ica} that given the validity of the first law of entanglement for boundary CFT states and the holographic dictionary for length metrics \eqref{hDictionary}, one can derive the linearised Einstein's equations around an empty AdS background. For AdS$_4$ with $\Lambda=-\frac{3}{L^2}$, the equations become:
\begin{align}
0=E_{\mn}&\equiv\frac{1}{2}\left(
    \nabla_\sigma \nabla_\mu h^\sigma_\nu
    +\nabla_\sigma \nabla_\nu h^\sigma_\mu
    -\nabla_\mu \nabla_\nu h - \square h_{\mn}
    \right)\nonumber
    \\
      & -\frac{1}{2}g_{\mn}\left(\nabla_\rho\nabla_\sigma h^{\rho\sigma}-\square h-\Lambda h\right)-\Lambda h_\mn.
    \label{LinearEinstein}
\end{align}
While the $\{t,t\}-$component of \eqref{LinearEinstein} is derived directly from the first law of entanglement, the $\{a,b\}-$components can be obtained by introducing a boost in the boundary directions. Finally the $\{z,\mu\}-$components imply the conservation and tracelessness of the metric perturbation $h_{ab}$:
\begin{equation}
    h^{a}_{a}=0, \,\partial^{a} h_{ab}=0 ~ \text{(all orders in $z$)}.
\end{equation}

In the first law of entanglement \eqref{first_law},
we use a power series expansion in $z$ for fluctuations on the LHS for finite $R$, while the RHS is given by just the leading order term due to the dictionary \eqref{hDictionary}. One then makes use of the equality between the two sides order by order in the expansion \footnote{The authors of \cite{Lashkari:2013koa} explicitly apply this method to derive linearised Einstein's equations}.

Coming to area metrics, now that we have the dictionary \eqref{wDictionary} for $w_{\mn\ab}
$, we can implement the same step by step analysis to find the corresponding equations of motion. At first, we again assume that the length-metric fluctuations $h_{\mn}$ are turned off. Since \eqref{wDictionary} is an identity about the rank two boundary EMT $T_{ab}$, we find the equations of motion to be rank two as well. Using this identity for $T_{tt}$, we find a second order differential equation to all orders in the power series expansion as (see appendix \ref{sec::appendixEOM} for a detailed analysis), 
\begin{align}
\label{Wequation}
  0=  &\frac{1}{z^4}\partial^{I}\partial^{J} W_{t{I}t{J}} +\partial_z^2 W_{tztz}+ \frac{2}{z} \left(\frac{1}{z^2}\partial^{I} W_{t{I}tz}+\partial_z W_{tztz}\right)\nonumber\\
    &+2\partial_z \left( \frac{1}{z^2}\partial^{I} W_{t{I}tz}\right).
\end{align}
This expression can be written more compactly by using the identity in \eqref{w_as_sum} that relates $w_{\mn \rho \lambda}$ with $W_{\mn \rho \lambda}$, yielding
\begin{equation}
\nabla^{\rho}\nabla^{\lambda} w_{t\rho t\lambda}=0,
\end{equation}
which we then extend to all boundary directions $\{a,b\}$ by demanding boost invariance. Finally, as a consistency check, if we naively extend these equations to $\{\mu,\nu\}$ (bulk) indices, we find that the $\{z,\mu\}$ equations of motions are satisfied if the conservation of the stress tensor holds for all orders in $z$, similar to the length metric case, 
\begin{equation}
     \partial^{a}\left(\eta^{ij} w_{a i b j}\right)=0 ~\text{(all orders in $z$).}
\end{equation}

Our proposal for covariant linearised equations for pure area metric fluctuations $w_{\mu\rho\nu\lambda}$ around empty $AdS_{4}$ is then
\begin{equation}
\nabla^{\rho}\nabla^{\lambda} w_{\mu\rho\nu\lambda}=0,
\end{equation}
which resembles the equation of motion of conformal gravity when comparing $w_{\mu\rho\nu\lambda}$ to a Weyl tensor  
\cite{Anastasiou_2016, maldacena2011einsteingravityconformalgravity, Manvelyan_2004, Dehghani_2008}, implying that the pure area metric like degrees of freedom couple conformally to the background spacetime. 
The boundary dual of these area metric fluctuations even in the absence of $h_{\mn}$ can be linked to an operator $\mathcal{O}_{w}$ with conformal dimension $h_{\mathcal{O}_w}=3$, which also happens to be the conformal dimension of the stress tensor in CFT$_3$. This supports our previous derivation of holographic dictionary relating the boundary stress tensor to the purely area metric fluctuations. 

Finally, turning on the length metric fluctuations $h_{\mn}$ again in addition to $w_{\mu\rho\nu\lambda}$, we can redo the same steps to derive the linearised equations of motion for a full area metric perturbation $a_{\mu\rho\nu\lambda}$ to be 
\begin{equation}
    E_\mn-\frac{1}{2}\nabla^{\rho}\nabla^{\lambda} w_{\mu\rho\nu\lambda}=0.
    \label{var_h_eq}
\end{equation}
This is the central result of our paper.

\section{Lagrangian formulation}

Having derived an equation of motion for the area metric, the question is now whether this follows from a Lagrangian. It is actually not clear to us whether an emergent gravity style derivation \textit{has} to arrive at an equation with a Lagrangian origin, but it would certainly be a nice feature to have. 

It is easy to see that \eqref{var_h_eq} follows from a Lagrangian of the form 
\begin{align}
    \mL=\mL_{EH}[h]-\frac{1}{2}h^{\ab}\nabla^{\mu} \nabla^{\nu} w_{\alpha \mu \beta \nu}+\mL_{w^2}[w], 
\label{L}
\end{align}
by taking the variation $\frac{\delta}{\delta h^\mn}$. Here, $\mL_{EH}[h]$ is the Lagrangian of linearised Einsteinian gravity, and $\mL_{w^2}[w]$ stands for terms independent of $h_\mn$ but quadratic in $w_{\mn\ab}$ or its derivatives. Partially following \cite{Borissova:2023yxs,Borissova:2025frj}, we consider the general ansatz
\begin{align}
    \mL_{w^2}=&-m^2 w_{\alpha \mu \beta \nu}w^{\alpha \mu \beta \nu}-K_1 
    (\nabla_\delta w_{\alpha \mu \beta \nu})(\nabla^\delta w^{\alpha \mu \beta \nu})
    \nonumber
    \\
    &-K_2 (\nabla^\alpha w_{\alpha \mu \beta \nu})
    (\nabla_\delta w^{\delta \mu \beta \nu}).
\label{Lsign}
\end{align}
Unfortunately, our results in the preceding sections does not directly constrain the parameters of the ansatz \eqref{Lsign} at all, however, there is something we can do about it. 
The holographic dictionary is usually derived from the holographic renormalisation procedure \cite{deHaro:2000vlm,Skenderis:2002wp}, and we anticipate that doing the same procedure (including identification of (non-)normalisable modes and proper boundary terms) for the action \eqref{Lsign} will in general yield a result inconsistent with \eqref{hwDictionary}. 
Demanding consistency with this result may then suffice to constrain the parameters of \eqref{Lsign} in non-trivial ways. 
The first step in this direction is ensuring that the equations of motion derived from \eqref{Lsign} allow for solutions with the asymptotic behaviour $w\sim z^{-1}$ which is required by \eqref{w_as_sum}. In the supplementary material, we identify
$    m^2=-\frac{6}{L^2}K_1$
as a promising ansatz. Further work in this direction is left to future publications, however.

\section{Introducing Lanczos potentials}

Taking the variation $\frac{\delta }{\delta w^{\alpha\beta\gamma\delta}}$ of \eqref{L}, the resulting equation of motion reads
\begin{align}
    \frac{1}{4}\delta^{(h)}_1 C_{\ab\gd}+...
    \label{C-constraint}
\end{align}
where the $+...$ stand for terms linear in $w$ (coming from $\mL_{w^2}$), and $\delta^{(h)}_1 C_{\ab\gd}$ represents the first order change in the bulk Weyl-tensor due to $h_{\mu\nu}$. This means that in the limit of ordinary Einsteinian \textit{geometry} ($w\rightarrow 0$), this equation yields a non-trivial constraint on $h_{\mn}$ which greatly reduces the space of solutions of \eqref{LinearEinstein}; we do not simply recover linearised Einsteinian \textit{gravity} in this limit.

A possible solution to this problem works as follows: Just like the Weyl tensor can always be written in terms of a Lanczos potential \cite{RevModPhys.34.379,Lanczos_Hist}, \textit{any} tensor of the form of $w_{\alpha\mu\beta\nu}$ can be written in terms of a potential \cite{Bampi1983}   
\begin{align}
\ell_{\ab\gamma}=-\ell_{\ba\gamma}, \  \ell_{[\ab\gamma]}=0,
\end{align}
see the supplementary material for the explicit expression. 
Taking the variation $\frac{\delta }{\delta \ell^{\ab\gamma}}$ of \eqref{L}, the term $\frac{-1}{2}h^{\ab}\nabla^{\mu} \nabla^{\nu} w_{\alpha \mu \beta \nu}$ (which is the origin of \eqref{C-constraint}), results in a 3rd-order differential expression for $h_{\mn}$ which reads 
\begin{align}
    F_{\ab\gamma}\equiv
    \frac{1}{3}g_{\alpha\gamma}\nabla_\beta E^\mu_\mu
    -\frac{1}{3}g_{\beta\gamma}\nabla_\alpha E^\mu_\mu
    -\nabla_\beta  E_{\alpha\gamma}
    +\nabla_\alpha E_{\beta\gamma}
    ,
\end{align}
and thus 
\begin{align}
    E_{\mn}=0\Rightarrow F_{\kappa\rho\sigma}=0. 
     \label{compatibility}
\end{align}
Building the bulk theory on $\ell$ as fundamental variable instead of $w$, we can hence re-obtain the linearised Einstein equations and their full set of solutions, without additional constraint, in the limit $w\rightarrow0$.

\section{Discussion}

To summarise, starting from the first law of entanglement and the Ryu-Takayanagi formula as first principles, we have systematically derived how area metrics can fit into the AdS/CFT correspondence in the linearised approximation. Our results establish the holographic dictionary for area metric fluctuations, how they couple to ordinary metric fluctuations, as well as what Lagrangian may give rise to such a bulk theory. We have also demonstrated how writing the purely area metric fluctuation in terms of a Lanczos potential yields a bulk model which reproduces Einstein's theory in the limit of ordinary metric geometry.

We expect these results to lay the groundwork for a more thorough investigation of holographic area metric gravity in the future. Obvious questions of interest may be the extension of our results to non-AdS backgrounds, beyond the leading order or beyond perturbation theory all-together, to cases including bulk matter fields, and to higher bulk dimensions. It will also be interesting to compare holographic results to others like e.g.~\cite{Schuller:2005ru,Giesel:2012vy,Ho:2015cza,Borissova:2023yxs,Borissova:2024cpx} and see whether there is a confluence of results coming from different theoretical frameworks, such as e.g.~string theory, LQG, and holography.

\begin{acknowledgments}
We wish to acknowledge many useful conversations with 
Shoaib Akhtar, 
Martin Ammon,
Micha\l\ P.~Heller,
Ted Jacobson,
Romuald Janik,
Cynthia Keeler, 
Elias Kiritsis, 
Nima Lashkari,
Juan Pedraza,
Zahra Raissi,
Andrew Rolph,
Rodrigo Andrade e Silva, 
Tadashi Takayanagi, 
Mark Van Raamsdonk, and
Leopoldo Pando Zayas.
\\

The work of both LC and MF as well as, up until 31.08.2025, AB was supported by the Polish National Science Centre (NCN) grant 2021/42/E/ST2/00234. AB acknowledges support from United Kingdom Research and Innovation (UKRI) under the UK government’s Horizon Europe guarantee (EP/Y00468X/1) (since 01.09.2025). LC acknowledges support from the Visibility and Mobility Module and the Research Support Module as part of the Excellence Initiative – Research University program at the Jagiellonian University, as well as from the European Union's Erasmus+ grant (Project No.: 2024-1-PL01-KA131-HED-000228791).
\\

Some of the calculations for this paper were performed using Wolfram Mathematica, and especially the packages diffgeo.m by Matthew Headrick \cite{diffgeo} and xAct \cite{xAct}. The data and codes that
support the findings of this article are openly available at \cite{UJ/2RMGIP_2025}. For the purpose of open
access, the authors have applied a Creative Commons Attribution (CC BY 4.0) licence to any
Author Accepted Manuscript version arising from this submission.

\end{acknowledgments}

\newpage

\appendix

\begin{widetext}

\section{Existence of dual EMT}

As explained in the main text, area metrics and length metrics are equivalent on the three dimensional boundary. They are related by the equations \cite{Ho:2015cza}
\begin{align}
G^{mnab}=g^{ma}g^{nb}-g^{na}g^{mb}
\label{AMfromMupup}
\end{align}
and 
\begin{align}
g^{ab}
=-\frac{1}{8}
\epsilon_{opq}\epsilon_{rst}
G^{aobr}G^{pqst}
=
-\frac{|g|}{8}
\vartheta_{opq}\vartheta_{rst}
G^{aobr}G^{pqst}.
\label{metricfromAMfinalIndicesUP2}
\end{align}
Here, we have picked a concrete sign, consistent with the Lorentzian case $g=\det[g_{ab}]<0$, and distinguished
\begin{align}
\text{Levi-Civita-\textit{symbol}:\ \ }    &\vartheta_{abc} 
\\
\text{Levi-Civita-\textit{tensor}:\ \ }    &\epsilon_{abc}= \sqrt{|g|}\vartheta_{abc} 
\end{align}
so that
\begin{align}
\frac{\delta \vartheta_{\ab\delta} }{\delta g_\mn}=0, \text{\ \ but\ \ }\frac{\delta \epsilon_{\ab\delta} }{\delta g_\mn}\neq 0.
\end{align}
We can also derive  
\begin{align}
\frac{8}{|g|}=    \sqrt{
\frac{2}{3}
\vartheta_{mnc}\vartheta_{def}\vartheta_{ghi}\vartheta_{jkl}
G^{mnde}G^{cgfj}G^{hikl}
}.
\label{|g|}
\end{align}
%
Plugging this into \eqref{metricfromAMfinalIndicesUP2} allows to express $g^{ab}$ as an expression that contains only $G^{wxyz}$ (no indices down) and the Levi-Civita-\textit{symbol} $\vartheta_{abc}$ which is independent of the metric as discussed above, thus providing an \textit{explicit} solution for the metric in terms of the area metric. 
Consequently, we can now derive the expressions for $\frac{\delta G^{mnrs}}{\delta g^{ab}}$ and $\frac{\delta g^{ab}}{\delta G^{mnrs}}$ which are necessary to translate between the EMT and the surface energy tensor. 
\\

From \eqref{AMfromMupup}, we easily obtain 
\begin{align}
&\frac{\delta G^{mnrs}}{\delta g^{ab}}
\nonumber
\\
&
=\frac{1}{2}\left(\delta_a^s\delta_b^n g^{mr}
+\delta_a^n\delta_b^s g^{mr}
-\delta_a^r\delta_b^n g^{ms}
-\delta_a^n\delta_b^r g^{ms}
-\delta_a^s\delta_b^m g^{nr}
-\delta_a^m\delta_b^s g^{nr}
+\delta_a^r\delta_b^m g^{ns}
+\delta_a^m\delta_b^r g^{ns}\right)
\\
&=\frac{1}{2}\overbracket[0.5pt][2pt]{\delta_a^s\delta_b^n g^{mr}}^{([mn][rs])}
\label{Gupgup}
\end{align}
with the notation
\begin{align}
\overbracket[0.5pt][2pt]{X^{abcd}}^{([ab][cd])}\equiv 
&X^{abcd}-X^{abdc}-X^{bacd}+X^{badc}+X^{cdab}-X^{cdba}-X^{dcab}+X^{dcba}, 
\label{notat}
\\
\underbracket[0.5pt][2pt]{X_{abcd}}_{([ab][cd])}\equiv 
&X_{abcd}-X_{abdc}-X_{bacd}+X_{badc}+X_{cdab}-X_{cdba}-X_{dcab}+X_{dcba}. 
\nonumber
\end{align}
From \eqref{metricfromAMfinalIndicesUP2}, we can derive 
\begin{align}
&\frac{\delta g^{ab}}{\delta G^{wxyz}}
=
-\frac{1}{8}
\epsilon_{opq}\epsilon_{rst}
\frac{\delta G^{aobr}}{\delta G^{wxyz}}G^{pqst}
-\frac{1}{8}
\epsilon_{opq}\epsilon_{rst}
G^{aobr}\frac{\delta G^{pqst}}{\delta G^{wxyz}}
-
\vartheta_{opq}\vartheta_{rst}
G^{aobr}G^{pqst}\frac{\delta |g|/8}{\delta G^{wxyz}}
\label{variationstep1}
\end{align}
The last term is the most challenging. Using \eqref{|g|},
\begin{align}
\frac{\delta |g|/8}{\delta G^{wxyz}}&=\frac{\delta }{\delta G^{wxyz}}\frac{1}{  \sqrt{
\frac{2}{3}
\vartheta_{mnc}\vartheta_{def}\vartheta_{ghi}\vartheta_{jkl}
G^{mnde}G^{cgfj}G^{hikl}
}}
\\
&=-\frac{1}{2(\sqrt{...})^3}\frac{\delta }{\delta G^{wxyz}}\left(
\frac{2}{3}
\vartheta_{mnc}\vartheta_{def}\vartheta_{ghi}\vartheta_{jkl}
G^{mnde}G^{cgfj}G^{hikl}
\right)
\\
&=-\frac{1}{2}\left(\frac{|g|}{8}\right)^3\left(\frac{16}{|g|^2}g_{wy}g_{xz}-\frac{16}{|g|^2}g_{wz}g_{xy}\right)
\\
&=\frac{|g|}{64}\left(g_{wz}g_{xy}-g_{wy}g_{xz}\right)
\end{align}
Taking everything together, \eqref{variationstep1} reads
\begin{align}
&\frac{\delta g^{ab}}{\delta G^{wxyz} }=
\label{gupGup}
\\
&\frac{1}{8}\left(
\delta_x^b \delta_z^a g_{wy}
+\delta_x^a \delta_z^b g_{wy}
-\delta_x^b \delta_y^a g_{wz}
-\delta_x^a \delta_y^b g_{wz}
-\delta_w^b \delta_z^a g_{xy}
-\delta_w^a \delta_z^b g_{xy}
+\delta_w^b \delta_y^a g_{xz}
+\delta_w^a \delta_y^b g_{xz}
+g_{wz}g_{xy}g^\ab-g_{wy}g_{xz}g^\ab
\right)
\nonumber
\\
&=\underbracket[0.5pt][2pt]{
\frac{1}{8}\delta_x^b\delta_z^a g_{wy}+\frac{1}{32}g_{wz}g_{xy}g^{ab}
}_{([wx][yz])}.
\nonumber
\end{align}
We can perform some consistency checks on these results by verifying that 
\begin{align}
\frac{\delta G^{mnrs}}{\delta g^{ab}}\frac{\delta g^{cd}}{\delta G^{mnrs} }=\frac{\delta g^{cd}}{\delta g^{ab}}\ \ \text{  and  }\ \ 
\frac{\delta G^{mnrs}}{\delta g^{ab}}\frac{\delta g^{ab}}{\delta G^{wxyz} }=\frac{\delta G^{mnrs}}{\delta G^{wxyz} }.
\end{align}
The latter depends on a dimensionally dependent identity valid in $d=3$. 
As stated in the main text, we can now define not only an energy momentum tensor $T_{ab}$, but also the \textit{surface energy tensor} $T_{mnrs}$ which encodes the same information. Specifically, 
using \eqref{Gupgup} we find that for a given $T_{mnrs}$
\begin{align}
T_{ab}=T_{mnrs}\frac{\delta G^{mnrs}  }{\delta g^{ab} }=4T_{a\ \ b m}^{\ m},
\label{Tab[Tmnrs]}
\end{align}
and likewise 
\begin{align}
 T_{wxyz}=   T_{ab}\frac{\delta g^{ab}}{\delta G^{wxyz} }=
 \frac{1}{4}g_{xz}T_{wy}
 -\frac{1}{4}g_{xy}T_{wz}
 -\frac{1}{4}g_{wz}T_{xy}
 +\frac{1}{4}g_{wy}T_{xz}
 +\frac{1}{8}g_{wz}g_{xy}T^c_c
 -\frac{1}{8}g_{wy}g_{xz}T^c_c
 \label{Tmnrs[Tab]}
\end{align}
In flat space (with $g_{ab}=\eta_{ab}$) for example, we find for the energy measured by a stationary observer 
\begin{align}
T_{00}=4 (T_{0101}+T_{0202}).
\label{T11}
\end{align}
So $T_{mnrs}$ might be interpreted as a tensor that measures the flow of energy within a given surface spanned by a bivector, and hence the energy flow in a certain direction is calculated by an equation like \eqref{T11} where we use bivectors with one direction along time and the other perpendicular to it, summing over orthonormal bivectors. Note that \eqref{Tab[Tmnrs]} implies 
\begin{align}
  T^a_a=4 T^{am}_{\ \ \ \ am}=2 G^{mnrs}T_{mnrs},
\end{align}
giving a very straightforward way to check for conformal invariance. Also, conservation of energy momentum implies
\begin{align}
\nabla_a T^{ab}=0\Leftrightarrow \nabla_a T^{a\ bm}_{\ m}=g_{nm}\nabla_a T^{anbm}=0.
\end{align}

\section{Holographic dictionary for the boundary EMT}
\label{sec::holodicti}

For the sake of clarity, let us first briefly discuss (following e.g.~\cite{Schuller:2005ru,Ho:2015cza}) the $0$-th order case, 
\begin{align}
    G_{\mn\ab}&= G^{(g)}_{\mn\ab}=g_{\mu\alpha}g_{\nu\beta}-g_{\nu\alpha}g_{\mu\beta}.
    \label{no-fluct}
\end{align}
By the defining property of area metrics \eqref{area_element}, the area of a codimension-2 surface $\Sigma$, parametrized by the embedding coordinates $X^\rho (\vec{x})$ in the $AdS_4$ background, is given by
\begin{align}
    A[\Sigma] \equiv \int_\Sigma d^2 x \ \sqrt{\frac{1}{4} \vartheta^{IJ} \ \hat{g}_{IJKL} (\vec{x}) \ \vartheta^{KL}},
\end{align}
where $\hat{g}_{IJKL}$ is the induced area metric on $\Sigma$:
\begin{align}
\hat{g}_{IJKL} (\vec{x})&= G_{\mn\ab} (X^\rho)\ \partial_{I} X^\mu \partial_{J} X^\nu \partial_{K} X^\alpha \partial_{L} X^\beta,
\end{align}
and $d^2 x = dx \wedge dy$. 
Interestingly, due to the symmetries of $\hat{g}_{IJKL}$ (which are identical to those of $G_{\mn\ab}$), only a single independent component, $\hat{g}_{1212}$, survives in two dimensions. 
Since the background is assumed to have a length metric description for now, this is precisely the determinant of the induced metric $\left(\hat{g}_{IJ} (\vec{x}) = g_{\mn} (X^\rho) \partial_{I} X^\mu \partial_{J} X^\nu\right)$ on $\Sigma$,
\begin{align}
    \det(\hat{g}_{IJ}) = \hat{g}_{11} \hat{g}_{22} - \hat{g}_{12} \hat{g}_{21}:= \hat{g} \equiv \hat{g}_{1212}.
\end{align}
Therefore, in the metric-induced case, the two prescriptions for computing the area of a surface coincide:
\begin{align}
    A[\Sigma] = \int_\Sigma d^2 x \ \sqrt{\frac{1}{4} \vartheta^{IJ} \ \hat{g}_{IJKL} (\vec{x}) \ \vartheta^{KL}} = \int_\Sigma d^2 x \ \sqrt{\hat{g}} .
\end{align}
However, adopting the area element prescription based on a general (not necessarily metric-induced) area metric tensor \eqref{area_element} allows one to explore more general fluctuations of the background spacetime.
\\

Going forward, following \eqref{AMfluctuation} and \eqref{fluctuationdecomp}, we will consider variations in the area of $\Sigma$ corresponding to cases when the length metric fluctuations of the bulk are still  turned off $(h_{\mn}=0)$ but purely area metric perturbations are allowed:
\begin{align}
    G_{\mn\ab}&= G^{(g)}_{\mn\ab} + w_{\mn\ab}\, ,\ \ (w\ll1).
    \label{w-fluct}
\end{align}
Due to the perturbation of the bulk, we find
\begin{align}
    A[\Sigma] &= \int_\Sigma d^2 x \ \sqrt{\hat{g} + \frac{1}{4} \vartheta^{IJ} w_{\mn\ab} (X^\rho)\ \partial_{I} X^\mu \partial_{J} X^\nu \partial_{K} X^\alpha \partial_{L} X^\beta \vartheta^{KL}}
    \\
    &= \int_\Sigma d^2 x \ \sqrt{\hat{g}} \left(1+\frac{1}{8} \frac{\vartheta^{IJ} w_{\mn\ab} (X^\rho)\ \partial_{I} X^\mu \partial_{J} X^\nu \partial_{K} X^\alpha \partial_{L} X^\beta \vartheta^{KL}}{\hat{g}}+...\right).
\end{align}
Therefore, for leading order area variations $\delta A$, we get
\begin{align}
    \delta A = \int_\Sigma d^2 x \  \left(\frac{1}{8} \frac{\vartheta^{IJ} w_{\mn\ab} (X^\rho)\ \partial_{I} X^\mu \partial_{J} X^\nu \partial_{K} X^\alpha \partial_{L} X^\beta \vartheta^{KL}}{\sqrt{\hat{g}}}\right).
\end{align}
Due to the properties of the Levi-Civita symbol $\vartheta^{IJ}$, the area variation simplifies to:
\begin{align}
    \delta A = \int_\Sigma d^2 x \  \left(\frac{1}{2} \frac{w_{\mn\ab} (X^\rho)\ \partial_1 X^\mu \partial_2 X^\nu \partial_1 X^\alpha \partial_2 X^\beta}{\sqrt{\hat{g}}}\right).
    \label{deltaA_simplified}
\end{align}
In the expression above, and in what follows, we mostly denote tensor components and derivatives using the indices $0, 1, 2, 3$. These should be understood as referring to the coordinates in the order $(t, x, y ,z)=(0, 1, 2, 3)$.

For the ball shaped spatial entangling subregions centred at $\vec{x}_{0}$ (where $\vec{x}_{0}=x^I_0=(x_0, y_0)$ with $I \in \{1, 2\}$) considered in this paper, the surface $\Sigma$ lies on a constant time slice $t = t_0$ and is parametrized as \cite{Ryu:2006ef}, 
\begin{align}
    z(x, y) = \sqrt{R^2-|\vec{x}-\vec{x}_0|^2},
    \label{embedding}
\end{align}
such that the only non-vanishing contributions to \eqref{deltaA_simplified} are:
\begin{align}
    \partial_1 X^1 &\equiv \partial_1 x=1, \, \partial_1 X^3 \equiv\partial_1 z=-\frac{(x-x_0)}{\sqrt{R^2-|\vec{x}-\vec{x}_0|^2}}, \\ \notag \partial_2 X^2 &\equiv\partial_2 y=1, \,\partial_2 X^3 \equiv\partial_2 z=-\frac{(y-y_0)}{\sqrt{R^2-|\vec{x}-\vec{x}_0|^2}}.
\end{align}
These simplifications and the symmetries of $w_{\mn \rho\lambda}$ amount to a handful of components contributing to $\delta A$,
\begin{align}
\label{varArea}
    \delta A =\int_{\Sigma} \frac{d^2x}{2L^2 R} z^3 \Bigg( &w_{1212}(z, t_0, \vec{x}) 
    + w_{1313}(z, t_0, \vec{x}) \frac{(y-y_0)^2}{R^2-|\vec{x}-\vec{x}_0|^2}
      + w_{3232}(z, t_0, \vec{x}) \frac{(x-x_0)^2}{R^2-|\vec{x}-\vec{x}_0|^2}
      \nonumber
      \\
      &-  2 w_{1213}(z, t_0, \vec{x}) \frac{(y-y_0)}{\sqrt{R^2-|\vec{x}-\vec{x}_0|^2}}
    -2w_{3212}(z, t_0, \vec{x}) \frac{(x-x_0)}{\sqrt{R^2-|\vec{x}-\vec{x}_0|^2}}
    \\
    &    +2w_{3213}(z, t_0, \vec{x}) \frac{(x-x_0)(y-y_0)}{R^2-|\vec{x}-\vec{x}_0|^2} \Bigg),
    \nonumber
\end{align}
where we have calculated the determinant of the induced metric to be $\hat{g}=L^4 R^2/z(x,y)^6$. In the $R \rightarrow 0$ limit, the integration region shrinks and the perturbation field $w_{\mn \ab}$ does not significantly vary across it. Therefore, under the integral sign, we can make the replacement
\begin{align}
    w_{\mn \ab} (z, t_0, \vec{x}) \approx w_{\mn \ab} (z, t_0, \vec{x}_0).
    \label{wInfinitesimal}
\end{align}
Moreover, after substituting \eqref{wInfinitesimal} into \eqref{varArea}, the last three terms in the integrand become odd functions of $x$ and $y$, and therefore vanish upon integration. This leaves us with the following expression
\begin{align}
\delta A= \int_{\Sigma} \frac{d^2 x}{2 L^2 R} z^3 \left( w_{1212} (z, t_0, \vec{x}_0)+ w_{1313}(z, t_0, \vec{x}_0) \frac{(y-y_0)^2}{R^2-|\vec{x}-\vec{x}_0|^2}+ w_{3232}(z, t_0, \vec{x}_0) \frac{(x-x_0)^2}{R^2-|\vec{x}-\vec{x}_0|^2}\right).
    \label{Areaintegrand_w_z}
\end{align}
In the next step, we write $w_{\mn \rho \lambda}$ as a power series in orders of $z$ centred around $z=0$. The leading order in this expansion will determine how $\delta A$ and hence $\delta S_A$ scales with $R$ as $R\rightarrow 0$. From \eqref{SEtensorEE}, we can tell that the correct scaling which yields a finite and non-zero expectation value for the boundary EMT is $\delta S_A\sim R^3$, and thus, with a bit of hindsight, we write   
\begin{align}
    w_{\mu \nu \rho \lambda} (z, t_0, \vec{x}) \equiv\frac{1}{z} W_{\mn \rho\lambda} (z, t_0, \vec{x}) = \frac{1}{z}\sum_{n=0} z^n W_{\mu \nu \rho \lambda}^{(n)} (t_0, \vec{x}).
    \label{z_expansion_w}
\end{align}
In this expression, the factor of $\frac{1}{z}$ highlights the leading order in $z-$ dependence of the pure area metric fluctuations. Upon inserting this power series in \eqref{Areaintegrand_w_z} and substituting the embedding function \eqref{embedding}, we find that the leading order term indeed scales as $R^3$ after the integration followed by $\mO(R^4), \mO(R^5), \cdots$ terms,
\begin{align}
  \delta A &=  \int_{\Sigma} \frac{d^2 x}{2 L^2 R} z^3 \left[\frac{1}{z} \left( W^{(0)}_{1212} (t_0, \vec{x}_0)
  + W^{(0)}_{1313}(t_0, \vec{x}_0)\frac{(y-y_0)^2}{R^2-|\vec{x}-\vec{x}_0|^2}
  + W^{(0)}_{3232}(t_0, \vec{x}_0)\frac{(x-x_0)^2 }{R^2-|\vec{x}-\vec{x}_0|^2}
  \right)+ \mO(z^0) + \cdots\right] \notag \\ 
    &= \frac{\pi R^3}{8 L^2} \left(2 W^{(0)}_{1212}(t_0, \vec{x}_0)+ W^{(0)}_{1313}(t_0, \vec{x}_0)+W^{(0)}_{3232}(t_0, \vec{x}_0)\right) + \mO(R^4) + \cdots .
\end{align}
With the help of the RT formula \cite{Ryu:2006bv, Ryu:2006ef}, we get
\begin{align}
    \delta S_A = \frac{\delta A}{4 G_N} = \frac{\pi R^3}{32 L^2 G_N} \left(2 W^{(0)}_{1212}(t_0, \vec{x}_0)+ W^{(0)}_{1313}(t_0, \vec{x}_0)+W^{(0)}_{3232}(t_0, \vec{x}_0)\right)+ \mO(R^4) + \cdots,
\end{align}
and from the expression for the holographic energy-momentum tensor \eqref{SEtensorEE}, we observe that the leading term in the power series of $w_{\mn \rho \lambda}$ provides the entry to the holographic dictionary for the dual EMT in the $R \rightarrow 0$ limit, 
\begin{align}
    \delta \langle T_{tt}(t_0, \vec{x}_0)\rangle&=\frac{2}{\pi^2} \lim_{R\rightarrow 0}\left(\frac{\pi}{32 L^2 G_N} \left(2 W^{(0)}_{1212}(t_0, \vec{x}_0)+ W^{(0)}_{1313}(t_0, \vec{x}_0)+W^{(0)}_{3232}(t_0, \vec{x}_0)\right)+ \mO(R)+\cdots\right) \notag \\ 
    &= \frac{1}{16 \pi L^2 G_N} \left(2 W^{(0)}_{1212}(t_0, \vec{x}_0)+ W^{(0)}_{1313}(t_0, \vec{x}_0)+W^{(0)}_{3232}(t_0, \vec{x}_0)\right).
\end{align}
We now recast this expression using the tracelessness of $w$, $g^{\mu\nu}w_{\alpha\mu\beta\nu} = 0$, which in the \Poincare-patch implies
\begin{align}
        w_{0101}&=w_{1212}+w_{1313},
        \nonumber
        \\
        w_{0202}&=w_{1212}+w_{3232},
                \label{TraceIdentities}
        \\
        w_{0303}&=w_{1313}+w_{3232},
        \nonumber
\end{align}
and thus
\begin{align}
    \delta \langle T_{tt}(t_0, \vec{x}_0)\rangle&= \frac{1}{16 \pi L^2 G_N} \left(W^{(0)}_{0101}(t_0, \vec{x}_0)+W^{(0)}_{0202}(t_0, \vec{x}_0)\right) =\frac{1}{16 \pi L^2 G_N} \eta^{ij} W^{(0)}_{0i0j}(t_0, \vec{x}_0).
\end{align}
Demanding invariance under Lorentz boosts along the boundary directions, we get
\begin{align}
    \delta \langle T_{ab}(t_0, \vec{x}_0)\rangle&= \frac{1}{16 \pi L^2 G_N} \eta^{ij} W^{(0)}_{aibj}(t_0, \vec{x}_0).
    \label{wDictionary_boostinv}
\end{align}
Now that we have derived the entry to the holographic dictionary for the boundary energy-momentum tensor, we require it to be traceless and conserved. Firstly, for tracelessness, we note that this property follows automatically from the tracelessness of $w$. Specifically, in the \Poincare AdS (where $g^{\ab} = z^2/L^2\text{diag} (-1,1,1,1)$ ),
\begin{align}
0\equiv g^{\beta\delta} w_{\ab\gd}=\frac{z^2}{L^2} w_{\alpha z\gamma z} + \frac{z^2}{L^2} \eta^{ij}w_{\alpha i\gamma j}
\label{trace condition 1}
\\
\Rightarrow 0\equiv \eta^{ij}w_{z iz j} \ \ \text{(by choosing $\alpha=\gamma=z$).}
\label{trace condition 2}
\end{align}
Taking the trace $g^{\alpha\gamma}$ of the expression above, we also find
\begin{align}
0 &\equiv  w_{z z z z} + w_{m z n z}\eta^{mn}+ \eta^{mn}\eta^{ij}w_{m in j}+\eta^{ij}w_{z i zj},
\\
\Rightarrow 0 &\equiv \eta^{mn}\eta^{ij}w_{m in j},
\label{trace condition w}
\end{align}
where we used \eqref{trace condition 2} and the index symmetry of $w_{\mn \rho \lambda}$ \eqref{symm1} to write \eqref{trace condition w}.
Now taking the trace of \eqref{wDictionary_boostinv}, we write
\begin{align}
& \delta \langle T^{a}_{\; a}\rangle = \eta^{ab}\delta \langle T_{ab}\rangle \sim \eta^{ab}\eta^{ij} W^{(0)}_{a i b j} =0.
\end{align}
Secondly, the conservation law of boundary EMT provides constraint equations for the $w-$fluctuations,
\begin{align}
    \partial^{a}\left(\delta \langle T_{ab}(t, \vec{x})\rangle\right)= 0 \Rightarrow  \partial^{a}\left( \eta^{ij} W^{(0)}_{aibj}(t, \vec{x})\right) =0.
\end{align}
We obtain the following three equations corresponding to the choices of the index $b$, with some simplifications due to the use of trace identities for $w$ \eqref{TraceIdentities}:
\begin{align}
    \text{For}\; b = 0:\,\,\,\,\,\,\,
    -\partial_{t} W^{(0)}_{0303}(t, \vec{x})&= \partial_x W^{(0)}_{0212}(t, \vec{x})+\partial_y W^{(0)}_{0121}(t, \vec{x}), \nonumber  \\     \text{For}\; b = 1: \qquad \partial_{t} W^{(0)}_{0212}(t, \vec{x})&= \partial_x W^{(0)}_{0202}(t, \vec{x})-\partial_y W^{(0)}_{0102}(t, \vec{x}), \\  
    \label{ConservationIdentities}
    \text{For}\; b = 2: \qquad \partial_{t} W^{(0)}_{0121}(t, \vec{x})&= -\partial_x W^{(0)}_{0102}(t, \vec{x})+\partial_y W^{(0)}_{0101}(t, \vec{x})\, \nonumber.
\end{align}

In this paper, we are dealing with linearised fluctuations around a fixed background spacetime and hence, all quantities, such as the holographic EMT, are linearised in perturbations. Therefore, when both length metric and pure area metric fluctuations are turned on, \textit{i.e.},
\begin{align}
    G_{\mn\ab}&= G^{(g)}_{\mn\ab} +2 h_{\alpha[\mu}g_{\nu]\beta}-2 h_{\beta[\mu}g_{\nu]\alpha} +w_{\mn\ab},
\end{align}
the holographic dictionary entry for the boundary EMT can be written as a sum of the known result from the study of length metric perturbations around $AdS$ \cite{deHaro:2015uta, Faulkner:2013ica, Lashkari:2013koa} and the result obtained in \eqref{wDictionary_boostinv},
\begin{align}
    &\delta \langle T_{ab}(t_0, \vec{x}_0)\rangle=\frac{1}{16\pi G_N} \left(3 H^{(0)}_{ab}(t_0, \vec{x}_0)+\frac{1}{L^2}\eta^{ij}W^{(0)}_{aibj}(t_0, \vec{x}_0)\right).
\end{align}

\section{Equations of motion}
\label{sec::appendixEOM}
In this section, we derive the equations of motion for area metric perturbations starting from the first law of entanglement, following the approach of \cite{Lashkari:2013koa}. Firstly, we consider the case when the length metric part of the full area metric is turned off \eqref{w-fluct}. For a ball-shaped region of finite radius $R$, we can calculate the variation in its entanglement entropy using \eqref{varArea} as follows,
\begin{align}
    \delta S_{A(R, 0, 0)} &= \frac{1}{8 L^2 G_N R} \int_{\Sigma} d^2 x\; z^3 \Bigg( w_{1212}(z, t_0, \vec{x})+ \frac{y^2}{z^2} w_{1313}(z, t_0, \vec{x}) + \frac{x^2}{z^2} w_{3232}(z, t_0, \vec{x})\notag\\& - \frac{2 y}{z} w_{1213}(z, t_0, \vec{x}) - \frac{2 x}{z} w_{3212}(z, t_0, \vec{x}) + \frac{2 xy}{z^2} w_{3213}(z, t_0, \vec{x})\Bigg).
\end{align}
Here, we have chosen the centre of the ball at $(t_0, 0, 0)$ without any loss of generality. 
Interestingly, in the finite ball analysis, all six integrands will contribute to the calculation of area because the region of integration is now large and finite, and the fluctuations $w_{\mn \rho \lambda}$ have an arbitrary dependence on $x$ and $y$. Upon inserting the power series expansion of $w_{\mn \rho\lambda}$  in orders of $z$ \eqref{z_expansion_w}, we obtain
\begin{align}
     \delta S_{A} &= \frac{1}{8 L^2 G_N R} \sum_{n=0} \int_{\Sigma} d^2 x\, \Bigg(z^{n+2}  W_{1212}^{(n)}(t_0, \vec{x})+ y^2 z^{n} W_{1313}^{(n)}(t_0, \vec{x}) + x^2 z^{n} W_{3232}^{(n)}(t_0, \vec{x}) \notag\\&- 2 y z^{n+1} W_{1213}^{(n)}(t_0, \vec{x}) - 2 x z^{n+1} W_{3212}^{(n)}(t_0, \vec{x}) + 2 xy z^{n} W_{3213}^{(n)}(t_0, \vec{x})\Bigg).
     \label{EOM_integrand_z_exp}
\end{align}
Furthermore, we expand the area metric fluctuations $W_{\mu \nu \rho \lambda}^{(n)}$ using the Taylor series around $\vec{x}= 0$,
\begin{align}
W_{\mu\nu\rho\lambda}^{(n)} (t_0, \vec{x}) = \sum_{m_x, m_y} \frac{1}{2 m_x!} \frac{1}{2 m_y!} x^{2m_x} y^{2m_y} \left[\partial_x^{2m_x} \partial_y^{2m_y} W_{\mu\nu\rho\lambda}^{(n)} \right](t_0, \vec{0}), \quad m_{x(y)} = 0,\frac{1}{2}, 1, \cdots ,
\label{Taylor-w}
\end{align}
where $\left[\partial_x^{2m_x} \partial_y^{2m_y} W_{\mu\nu\rho\lambda}^{(n)}\right] (t_0, \vec{0})$ implies derivatives of $W_{\mu\nu\rho\lambda}^{(n)}(t_0,\vec{x})$ evaluated at $\vec{x} = (0,0)$. We insert this series in $\eqref{EOM_integrand_z_exp}$ along with the embedding function $z(x, y)$ \eqref{embedding}, to write $\delta S_{A}$ as
\begin{align}
    \delta S_{A} &= \frac{1}{8 L^2 G_N R} \sum_{n=0} \sum_{m_x, m_y} \frac{1}{(2m_x)!(2m_y)!} \int_{\Sigma} d^2 x\, \left[
    \begin{aligned}
        & (R^2-x^2-y^2)^{\frac{n+2}{2}} x^{2m_x} y^{2m_y} \left[\partial_x^{2m_x}\partial_y^{2m_y} W_{1212}^{(n)}\right] (t_0, \vec{0})
        \\&+ (R^2-x^2-y^2)^{\frac{n}{2}} x^{2m_x} y^{2m_y+2} \left[\partial_x^{2m_x}\partial_y^{2m_y} W_{1313}^{(n)}) \right](t_0, \vec{0})
        \\&+(R^2-x^2-y^2)^{\frac{n}{2}} x^{2m_x+2} y^{2m_y} \left[\partial_x^{2m_x}\partial_y^{2m_y} W_{3232}^{(n)} \right](t_0, \vec{0})
        \\&-2 (R^2-x^2-y^2)^{\frac{n+1}{2}} x^{2m_x} y^{2m_y+1} \left[\partial_x^{2m_x}\partial_y^{2m_y} W_{1213}^{(n)}\right] (t_0, \vec{0})
        \\&-2(R^2-x^2-y^2)^{\frac{n+1}{2}} x^{2m_x+1} y^{2m_y} \left[\partial_x^{2m_x}\partial_y^{2m_y} W_{3212}^{(n)} \right](t_0, \vec{0})
        \\&+2(R^2-x^2-y^2)^{\frac{n}{2}} x^{2m_x+1} y^{2m_y+1} \left[\partial_x^{2m_x}\partial_y^{2m_y} W_{3213}^{(n)}\right](t_0, \vec{0})
    \end{aligned}\right].
\end{align}
To perform this integration, we need to make use of the following integral identities \cite{Lashkari:2013koa},
\begin{align}
\label{basic integrals}
    \int_{\Sigma} d^2 x\, (R^2 -x^2-y^2)^{\frac{n}{2}} x^{2 m_x} y^{2 m_y} &= R^{n +2 m_x +2 m_y + 2} I_{n, m_x, m_y}, \quad m_{x(y)}  = 0, 1, 2,\cdots, \notag \\
    \int_{\Sigma} d^2 x\, (R^2 -x^2-y^2)^{\frac{n+1}{2}} x^{2 m_{x}+1} y^{2 m_{y}} &= R^{n +2 m_x +2 m_y + 4} I_{n+1, m_x+\frac{1}{2}, m_y}, \quad m_{x}  = \frac{1}{2}, \frac{3}{2}, \frac{5}{2},\cdots, \quad m_{y}  = 0, 1, 2,\cdots, \notag \\
    \int_{\Sigma} d^2 x\, (R^2 -x^2-y^2)^{\frac{n}{2}} x^{2 m_x+1} y^{2 m_y+1} &= R^{n +2 m_x +2 m_y + 4} I_{n, m_x +\frac{1}{2}, m_y+\frac{1}{2}}, \quad m_{x(y)}=\frac{1}{2}, \frac{3}{2}, \frac{5}{2},\cdots ,
\end{align}
where 
\begin{align}
    I_{n, m_x, m_y} = \frac{\Gamma\left(\frac{n}{2}+1\right) \Gamma\left(m_x+\frac{1}{2}\right)\Gamma\left(m_y+\frac{1}{2}\right)}{\Gamma\left(\frac{n}{2}+m_x + m_y +2\right)}.
\end{align}
Upon integrating using the identities of \eqref{basic integrals}, we obtain $\delta S_{A}$ in the orders of $R$,
\begingroup
\small
\begin{align}
     & \frac{1}{8 L^2 G_N R} \sum_{n=0} \sum_{m_x, m_y} \frac{R^{n+2m_x+2m_y+4}}{(2m_x)!(2m_y)!} \left[
    \begin{aligned}
        &I_{n+2, m_x, m_y} \left[\partial_x^{2m_x}\partial_y^{2m_y}W_{1212}^{(n)}\right](t_0, \vec{0})+ I_{n, m_x, m_y+1} \left[\partial_x^{2m_x}\partial_y^{2m_y}W_{1313}^{(n)}\right](t_0, \vec{0}) \\&+ I_{n, m_x+1, m_y} \left[\partial_x^{2m_x}\partial_y^{2m_y}W_{3232}^{(n)}\right](t_0, \vec{0})- 2 I_{n+1, m_x, m_y+\frac{1}{2}} \left[\partial_x^{2m_x}\partial_y^{2m_y}W_{1213}^{(n)}\right] (t_0, \vec{0})\\&- 2 I_{n+1, m_x+\frac{1}{2}, m_y} \left[\partial_x^{2m_x}\partial_y^{2m_y}W_{3212}^{(n)}\right](t_0, \vec{0})+ 2 I_{n, m_x+\frac{1}{2}, m_y+\frac{1}{2}} \left[\partial_x^{2m_x}\partial_y^{2m_y}W_{3213}^{(n)}\right](t_0, \vec{0})  
    \end{aligned}\right].
\end{align}
\endgroup
We have not written the allowed values of $m_x$ and $m_y$ in the summation for the sake of compactness of the above expression, but it should be understood that when $m_{x(y)}$ or $m_{x(y)+1}$ appear in $I.,.,.$, then the allowed values are $m_{x(y)} = 0, 1, 2, \cdots$, and when $m_{x(y)+\frac{1}{2}}$ appear then the possible values are $m_{x(y)} = \frac{1}{2}, \frac{3}{2}, \frac{5}{2}, \cdots$\, for non-vanishing integrands. 

We make a change of variable $m_{x(y)} = m'_{x(y)} +\frac{1}{2}$ in the last three terms of the above equation so that the summation runs over $m_{x(y)} = 0, 1, 2, \cdots$ and while doing so, absorb the incoming extra factors of $R$ in the possible values of $n$ such that,
\begin{align}
    \sum_{n=0} \;\sum_{m_x} \sum_{m_y= 1/2, \cdots} &\frac{R^{n+ 2 m_x +2 m_y +4}}{(2 m_x)! (2m_y)!} I_{n+1, m_x, m_y+\frac{1}{2}} \left[\partial_x^{2m_x} \partial_y^{2m_y}W_{1213}^{(n)}\right](t_0, \vec{0}) 
    \notag \\&= \sum_{n=1}\; \sum_{m_x, m'_y= 0, 1, \cdots} \frac{R^{n+ 2 m_x +2 m'_y +4}}{(2 m_x)! (2m'_y+1)!} I_{n, m_x, m'_y+1} \left[\partial_x^{2m_x} \partial_y^{2m'_y}\partial_y W_{1213}^{(n-1)}\right](t_0, \vec{0}), 
    \\[10pt]
     \sum_{n=0} \;\sum_{m_y}\sum_{m_x= 1/2, \cdots} &\frac{R^{n+ 2 m_x +2 m_y +4}}{(2 m_x)! (2m_y)!} I_{n, m_x+\frac{1}{2}, m_y} \left[\partial_x^{2m_x} \partial_y^{2m_y}W_{3212}^{(n)}\right](t_0, \vec{0}) \notag \\&= \sum_{n=1}\; \sum_{m'_x, m_y= 0, 1, \cdots} \frac{R^{n+ 2 m'_x +2 m_y +4}}{(2 m'_x+1)! (2m_y)!} I_{n, m'_x+1, m_y} \left[\partial_x^{2m'_x} \partial_y^{2m_y}\partial_x W_{3212}^{(n-1)}\right](t_0, \vec{0}), 
     \\[10pt]
     \sum_{n=0} \;\sum_{m_x, m_y= 1/2, \cdots} &\frac{R^{n+ 2 m_x +2 m_y +4}}{(2 m_x)! (2m_y)!} I_{n, m_x+\frac{1}{2}, m_y+\frac{1}{2}} \left[\partial_x^{2m_x} \partial_y^{2m_y}W_{3213}^{(n)}\right](t_0, \vec{0}) \notag \\&= \sum_{n=2}\; \sum_{m'_x, m'_y= 0, 1, \cdots} \frac{R^{n+ 2 m'_x +2 m'_y +4}}{(2 m'_x+1)! (2m'_y+1)!} I_{n-2, m'_x+1, m'_y+1} \left[\partial_x^{2m'_x} \partial_y^{2m'_y}\partial_x \partial_y W_{3213}^{(n-2)}\right](t_0, \vec{0}).
\end{align}
We incorporate all these modifications to finally write down the LHS of the first law of entanglement, $\delta S_{A}$, as
\begingroup
\small
\begin{align}
    \delta S_A &=\frac{1}{8 L^2 G_N R}\sum_{m_x, m_y} \frac{R^{2m_x+2m_y+4}}{(2m_x)!(2m_y)!}\times
    \notag\\& \left[
    \begin{aligned}
        &\sum_{n=0} R^{n}\left(I_{n+2, m_x, m_y} \left[\partial_x^{2m_x}\partial_y^{2m_y}W_{1212}^{(n)}\right](t_0, \vec{0})+ I_{n, m_x, m_y+1} \left[\partial_x^{2m_x}\partial_y^{2m_y}W_{1313}^{(n)}\right](t_0, \vec{0})+ I_{n, m_x+1, m_y}\left[\partial_x^{2m_x}\partial_y^{2m_y} W_{3232}^{(n)}\right](t_0, \vec{0})\right)
        \\&- \sum_{n=1} R^{n} \left(\frac{2}{(2m_y+1)} I_{n, m_x, m_y+1} \left[\partial_x^{2m_x}\partial_y^{2m_y+1} W_{1213}^{(n-1)}\right](t_0, \vec{0})+ \frac{2}{(2m_x+1)} I_{n, m_x+1, m_y} \left[\partial_x^{2m_x+1}\partial_y^{2m_y} W_{3212}^{(n-1)}\right](t_0, \vec{0})\right)
        \\& + \sum_{n=2} R^{n}
        \left(\frac{2}{(2m_x+1)(2m_y+1)} I_{n-2, m_x+1, m_y+1} \left[\partial_x^{2m_x+1}\partial_y^{2m_y+1}W_{3213}^{(n-2)}\right](t_0, \vec{0})\right)
    \end{aligned}\right].
\end{align}
\endgroup
Now, let us turn our attention to the RHS of the first law of entanglement \eqref{first_law}. In the previous subsection, we have derived the holographic dictionary for the boundary stress-energy tensor which reads,
\begin{align}
    \delta \langle T_{tt}(t_0, \vec{x}_0)\rangle = \frac{1}{16 \pi L^2 G_N } \left(2 W_{1212}^{(0)}(t_0, \vec{x}_0) + W_{1313}^{(0)}(t_0, \vec{x}_0)+W_{3232}^{(0)}(t_0, \vec{x}_0)\right).
\end{align}
We insert this dictionary in \eqref{defdeltaH} and obtain an expression for $\delta \langle H_A \rangle$ as,
\begin{align}
    \delta \langle H_A \rangle = \frac{1}{16 L^2 G_N } \int_{A(R, 0, 0)} d^2 x \, \frac{R^2-|\vec{x}|^2}{R} \left(2 W_{1212}^{(0)} (t_0, \vec{x}) + W_{1313}^{(0)}(t_0, \vec{x})+W_{3232}^{(0)}(t_0, \vec{x})\right).
\end{align}
Now, we expand the area metric components using the Taylor series expansion around $\vec{x} =\vec{0}$ \eqref{Taylor-w},
\begin{align}
     \delta \langle H_A \rangle = &\frac{1}{16 L^2 G_N }  \int_{A(R, 0, 0)} d^2 x \, \frac{R^2-|\vec{x}|^2}{R} \times 
     \\&\sum_{m_x, m_y} \frac{x^{2m_x}}{2 m_x!} \frac{y^{2m_y} }{2 m_y!} \Big(2\left[\partial_x^{2m_x} \partial_y^{2m_y} W_{1212}^{(0)}\right] (t_0, \vec{0}) + \left[\partial_x^{2m_x} \partial_y^{2m_y}W_{1313}^{(0)}\right](t_0, \vec{0})+\left[\partial_x^{2m_x} \partial_y^{2m_y}W_{3232}^{(0)}\right](t_0, \vec{0})\Big),
     \nonumber
\end{align}
where the summation runs over $m_{x(y)} = 0, 1, 2, \cdots$.
It is important to note that the $W_{\mn \rho \lambda}^{(0)}$ are not functions of the bulk radial coordinate $z$, so we do not have a summation over the parameter $n$ here.  Finally, we integrate the above expression for $\delta \langle H_A \rangle$ by utilizing the identities of \eqref{basic integrals}:
\begin{align}
    \delta \langle H_A \rangle& = \frac{1}{16 L^2 G_N R} \sum_{m_x, m_y} \frac{R^{2m_x+2m_y+4}}{(2 m_x)!(2 m_y)!} I_{2, m_x, m_y}\times \nonumber\\& \left(2\left[\partial_x^{2m_x} \partial_y^{2m_y} W_{1212}^{(0)}\right] (t_0, \vec{0}) + \left[\partial_x^{2m_x} \partial_y^{2m_y}W_{1313}^{(0)}\right](t_0, \vec{0})+\left[\partial_x^{2m_x} \partial_y^{2m_y}W_{3232}^{(0)}\right](t_0, \vec{0})\right).
\end{align}
To obtain a version of linearised EOM for area metric perturbations, we compare orders of $R$ on both sides of $\delta S_{A} =\delta \langle H_A \rangle$. The outcome at different orders of $R$ is listed below,
\begin{align}
    &\text{at} \; \mathcal{O}(R^3): \delta S_A-\delta \langle H_A \rangle = 0 \implies 0=0, \notag \\    &\text{at} \; \mathcal{O}(R^4): \delta S_A-\delta \langle H_A \rangle = 0 \implies 3 W_{1212}^{(1)}+W_{1313}^{(1)}+W_{3232}^{(1)} \equiv 2 W_{1212}^{(1)}=2\partial_y W_{1213}^{(0)}+2\partial_x W_{3212}^{(0)}, \notag \\ 
    &\text{at} \; \mathcal{O}(R^5): \delta S_A-\delta \langle H_A \rangle = 0 \implies 2 \partial_x \partial_y W_{3213}^{(0)} = -\partial_x^2 W_{3232}^{(0)}-\partial_y^2 W_{1313}^{(0)} + 4 (\partial_x W_{3212}^{(1)}+\partial_y W_{1213}^{(1)})-6 W_{1212}^{(2)}, \notag \\ 
    &\text{at} \; \mathcal{O}(R^6): \delta S_A-\delta \langle H_A \rangle = 0 \implies 2 \partial_x \partial_y W_{3213}^{(1)} = -\partial_x^2 W_{3232}^{(1)}-\partial_y^2 W_{1313}^{(1)} + 6 (\partial_x W_{3212}^{(2)}+\partial_y W_{1213}^{(2)})-12 W_{1212}^{(3)},\notag \\ 
    &\text{at} \; \mathcal{O}(R^7): \delta S_A-\delta \langle H_A \rangle = 0 \implies 2 \partial_x \partial_y W_{3213}^{(2)} = -\partial_x^2 W_{3232}^{(2)}-\partial_y^2 W_{1313}^{(2)} + 8 (\partial_x W_{3212}^{(3)}+\partial_y W_{1213}^{(3)})-20 W_{1212}^{(4)}, \notag \\
    &\vdots \notag \\ 
    &\text{in general}, \notag \\
    &\delta S_A-\delta \langle H_A \rangle = 0 \implies 2 \partial_x \partial_y W_{3213}^{(n-2)} = -\partial_x^2 W_{3232}^{(n-2)}-\partial_y^2 W_{1313}^{(n-2)} + 2 n (\partial_x W_{3212}^{(n-1)}+\partial_y W_{1213}^{(n-1)})-n(n+1)W_{1212}^{(n)}
\end{align}
for $n\geq 2$. 
Using the tracelessness of $w_{\mn \rho \lambda}$ \eqref{TraceIdentities}
, we re-write the above equation as,
\begin{align}
\label{area EOM order n}
    2 \partial_x \partial_y W_{0102}^{(n-2)} = -\left(\partial_x^2 W_{0101}^{(n-2)}+\partial_y^2 W_{0202}^{(n-2)} + 2n (\partial_x W_{0103}^{(n-1)}+\partial_y W_{0203}^{(n-1)})+n(n+1)W_{0303}^{(n)}\right).
\end{align}
This recursive relation can be obtained by inserting the power series expansion of $W_{\mn \rho \lambda}$ \eqref{z_expansion_w} in the following general equation:
\begin{equation}
\label{areaEOMfromentanglement}
    -2 \partial_x \partial_y W_{0102} = \partial_x^2 W_{0101}+\partial_y^2 W_{0202} + \frac{2}{z} \partial_z\left(z\left(\partial_x W_{0103}+\partial_y W_{0203}\right)\right)+\frac{1}{z^2}\partial_z (z^2 \partial_z W_{0303}).
\end{equation}
Upon further solving,
\begin{align}
    -2 (\partial_x \partial_y W_{0102}+ \partial_x \partial_z W_{0103}+\partial_y \partial_z W_{0203})= \partial_x^2 W_{0101}+\partial_y^2 W_{0202} +\partial_z^2 W_{0303}+ \frac{2}{z} \left(\partial_x W_{0103}+\partial_y W_{0203}+\partial_z W_{0303}\right).
\end{align}
Now, it is straightforward to verify that the above equation can be written in a compact form as
\begin{align}\label{00areametric}
    \partial^{\mu} \partial^{\nu} W_{0 \mu 0 \nu} = 0, \quad \text{where} \;\mu, \nu = 0, 1, 2, 3.
\end{align}
Hence, we have derived the $\{t, t\}-$component of the linearised EOM for pure area metric fluctuations. Furthermore, to preserve boost invariance, it is necessary to extend this equation for all the boundary directions instead of just one component,
\begin{align}\label{mn_areametric}
    \partial^{\mu} \partial^{\nu} W_{m \mu n \nu} = 0, \quad \text{where} \;m, n = 0, 1, 2.
\end{align}
Since the background spacetime is curved, we should expect the EOM for fluctuations to contain covariant derivatives, not just simple partial derivatives. We can check that the following expression, written in terms of $w$, is equivalent to \eqref{mn_areametric}:
\begin{align}\label{cov_mn_areametric}
    \nabla^{\mu} \nabla^{\nu} w_{m \mu n \nu} = 0.
\end{align}
In the final step, our goal is to write down an EOM for area metric fluctuations which is valid for all bulk dimensions. From our analysis so far, the candidate equation should take the form:
\begin{align}\label{areametric_EOM}
    \nabla^{\mu} \nabla^{\nu} w_{\alpha \mu \beta \nu} = 0, \quad \text{where} \;\alpha, \beta, \mu, \nu = 0, 1, 2, 3.
\end{align}

Once again, if both the length-metric and pure area metric fluctuations are turned on \eqref{fluctuationdecomp}, the full EOM linearised in perturbations take the form of a linear combination of the previously obtained results,
\begin{align}
    E_\mn-\frac{1}{2}\nabla^{\rho}\nabla^{\lambda} w_{\mu\rho\nu\lambda}=0,
\end{align}
where $E_{\mn}$ is the contribution \eqref{LinearEinstein} from length-metric  fluctuations $h_{\mn}$ that corresponds to linearised Einstein's equations. 
The relative factor of $-\frac{1}{2}$ arises from a careful comparison of our calculations in this section leading up to \eqref{areametric_EOM} and the derivation of $E_\mn=0$ in \cite{Blanco:2013joa,Lashkari:2013koa,Faulkner:2013ica}.

\section{Lagrangian formulation}
\label{sec::Lagrangian_EOM_w}

\subsection{Role of tracelessness and the need for Lagrange multipliers}
\label{LM_for_w}

Before deriving area metric equations by taking the variation of the action  with respect to $w_{\ab\mn}$, it will be instructive to consider the following simple toymodel: Let's say we have a tensor field $\phi_\mn$ which is required to be symmetric \textit{and traceless}, and described by a Lagrangian $\mL_\phi$ containing at most first derivatives of $\phi$. Setting the variation of the action to zero, we find (ignoring boundary terms) 
\begin{align}
 0\equiv   \delta \mS &= \int dV \delta \mL_\phi
 \label{variation}
 \\
 &= \int dV \left(\frac{\partial \mL_\phi}{\partial \phi_{\ab}}-\partial_\gamma\frac{\partial \mL_\phi}{\partial (\partial_\gamma\phi_{\ab})}\right)\delta \phi^\ab.
\end{align}
From this we can read off what we will from now on call the "naive" Euler-Lagrange equation
\begin{align}
0=\frac{\partial \mL_\phi}{\partial \phi_{\ab}}-\partial_\gamma\frac{\partial \mL_\phi}{\partial (\partial_\gamma\phi_{\ab})} \equiv X_\ab. 
\label{EOMnaive}
\end{align}
This naive equation is however \textit{not} a necessary condition for \eqref{variation}:  As we impose $\phi$ to be traceless, we should only consider variations of the field that also satisfy $\delta \phi^\mu_\mu=0$, and so a necessary condition for \eqref{variation} is merely 
\begin{align}
    X_\ab =Y g_\ab
    \label{XY}
\end{align}
with any arbitrary function $Y$. To treat this issue more formally, we add a Langrange-multiplier $\lambda$ to the Lagrangian enforcing the trace condition explicitly:
\begin{align}
\mL'_\phi=\mL_\phi+\lambda g_\mn \phi^\mn
\end{align}
Now the Euler-Lagrange equation for the field $\lambda$ enforces the holonomic constraint $\phi^\mu_\mu=0$, and the Euler-Lagrange equation for $\phi_\ab$ reads
\begin{align}
X_\ab+\lambda g_{\ab}= 0
\label{LagrangeEOM}
\end{align}
which is equivalent to our heuristic result \eqref{XY}. Solving \eqref{LagrangeEOM} for $\lambda$, we simply take the trace to find 
$
4\lambda \equiv -X^\mu_\mu,
$
and \eqref{LagrangeEOM} takes the form
\begin{align}
  X_\ab- \frac{X^\rho_\rho}{4} g_\ab= 0. 
 \label{EOMtraceless}
\end{align}
In essence, we have found that the true physical equation of motion \eqref{EOMtraceless} is only the traceless part of the naive equation of motion \eqref{EOMnaive}. 
\\

With this in mind, we now leave the above toymodel behind and return to considering the purely area metric fluctuation $w_{\ab\mn}$, its action and its equations of motion. Here, we actually have to enforce two non-trivial constraints, namely tracelessness and cyclicity. To do so, we add two Lagrangian multiplier field terms to the Lagrangian \eqref{L},
\begin{align}
\mL'=\mL+\lambda^\ab g^\mn w_{\alpha\mu\beta\nu} + \kappa \epsilon^{\alpha\mu\beta\nu}w_{\alpha\mu\beta\nu},
\end{align}
with $\lambda^{\ab}=\lambda^{\ba}$. Variation with respect to $\kappa$ enforces the cyclicity condition (second part of \eqref{TraceW}), while variation with respect to $\lambda^\ab$ enforces tracelessness (first part of \eqref{TraceW}). Variation  with respect to $w_{\alpha\mu\beta\nu}$ then gives equations of motion of the form 
\begin{align}
    0\equiv X_{\alpha\mu\beta\nu}+\frac{1}{4}g_{\mu\nu}\lambda_{\alpha\beta}
    -\frac{1}{4}g_{\mu\beta}\lambda_{\alpha\nu}
    -\frac{1}{4}g_{\alpha\nu}\lambda_{\mu\beta}
    +\frac{1}{4}g_{\alpha\beta}\lambda_{\mu\nu}+\epsilon_{\alpha\mu\beta\nu}\kappa
    \label{lambda_ab_4indexeq}
\end{align}
where $X_{\alpha\mu\beta\nu}$ stands again for the "naive" Euler-Lagrange equations derived from $\mL$,
\begin{align}
    X_{\alpha\mu\beta\nu}\equiv \frac{\partial \mL}{\partial w_{\alpha\mu\beta\nu}}-\partial_\gamma\frac{\partial \mL}{\partial (\partial_\gamma w_{\alpha\mu\beta\nu})}. 
\end{align}
As in the toy model above, we now have to solve for $\lambda_{\ab}$ and $\kappa$. Taking both traces of \eqref{lambda_ab_4indexeq}, thanks to $\epsilon_{abcd}g^{bd}=0$, we get
\begin{align}
    0\equiv X^{\alpha\mu}_{\ \ \alpha\mu}+\frac{3}{2}\lambda^\alpha_{\alpha}
    \label{lambda_ab_0indexeq}.
\end{align}
Taking one trace of \eqref{lambda_ab_4indexeq} and using \eqref{lambda_ab_0indexeq}, we find 
\begin{align}
    0\equiv 2X^{\ \alpha}_{\mu\ \nu \alpha}
    -\frac{1}{3}g_{\mu\nu}X^{\alpha \beta}_{\ \ \alpha \beta}
    +\lambda_{\mu\nu}
    \label{lambda_ab_2indexeq}
\end{align}
which can be solved for $\lambda_{\mu\nu}$. Likewise, contracting \eqref{lambda_ab_4indexeq} with $\epsilon^{\alpha\mu\beta\nu}$ we get
\begin{align}
    0\equiv -\epsilon^{\alpha\mu\beta\nu}X_{\alpha\mu\beta\nu}+24\kappa
\end{align}
which can be solved for $\kappa$. With these expressions at hand, the equation of motion \eqref{lambda_ab_4indexeq} finally reads
\begin{align}
    0\equiv &X_{\alpha\mu\beta\nu}
    \nonumber
    \\
    &-\frac{1}{2}g_{\mu\nu} X^{\ \gamma}_{\alpha\ \beta\gamma}
    +\frac{1}{2}g_{\mu\beta} X^{\ \gamma}_{\alpha\ \nu\gamma}
    +\frac{1}{2}g_{\alpha\nu} X^{\ \gamma}_{\mu\ \beta\gamma}
    -\frac{1}{2}g_{\alpha\beta} X^{\ \gamma}_{\mu\ \nu\gamma}
    -\frac{1}{6}g_{\alpha\nu}g_{\mu\beta} X^{\gamma\delta}_{\ \ \gamma\delta}
    +\frac{1}{6}g_{\alpha\beta}g_{\mu\nu} X^{\gamma\delta}_{\ \ \gamma\delta}
    \label{physical_EOMs}
    \\
    &+\frac{1}{24}\epsilon_{\alpha\mu\beta\nu}\epsilon^{\gamma\delta\rho\sigma}X_{\gamma\delta\rho\sigma}.
    \nonumber
\end{align}
This expression relates to the naive expression $X_{\alpha\mu\beta\nu}$ like the Weyl tensor relates to the Riemann tensor, i.e.~it is only the traceless part of the original expression (in addition to the fact that any potential non-cyclic part of $X_{\alpha\mu\beta\nu}$ is explicitly removed).

\subsection{Equations of motion}

Following equations \eqref{L} and \eqref{Lsign} in the main text as well as the results of the previous subsection, we now consider a bulk Lagrangian of the full form 
\begin{align}
    \mL=&\mL_{EH}[h]+\mL_{h\text{-}w}[h,w]+\mL_{w^2}[w]+\mL_{LM}[w,\lambda,\kappa]
    \label{Lfull}
\end{align}
with 
\begin{align}
\mL_{EH}[h]&=
-\frac{1}{4}\tilde{h}^\mn\Box h_{\mn}
-\frac{1}{2}\nabla^\rho\tilde{h}_{\rho\mu}\nabla^\sigma\tilde{h}_\sigma^\mu
+\frac{\Lambda}{6}\tilde{h}^\mn h_\mn
+\frac{\Lambda}{6}h^2
 \, \left(\text{with}\,\, \tilde{h}_\mn=h_\mn-\frac{1}{2}g_{\mn}h\right),
\\
\mL_{h\text{-}w}[h,w]&=-\frac{1}{2}h^{\ab}\nabla^{\mu} \nabla^{\nu} w_{\alpha \mu \beta \nu},
\\
\mL_{w^2}[w]&=-m^2 w_{\alpha \mu \beta \nu}w^{\alpha \mu \beta \nu}-K_1 
    (\nabla_\delta w_{\alpha \mu \beta \nu})(\nabla^\delta w^{\alpha \mu \beta \nu})-K_2 (\nabla^\alpha w_{\alpha \mu \beta \nu})
    (\nabla_\delta w^{\delta \mu \beta \nu}),
    \label{Lfull_w2}
\\
\mL_{LM}[w,\lambda,\kappa]&= \lambda^\ab g^\mn w_{\alpha\mu\beta\nu} + \kappa \epsilon^{\alpha\mu\beta\nu}w_{\alpha\mu\beta\nu}.
\end{align}
The terms in the Lagrangian are, in this order, the Einstein-Hilbert term linearised around the AdS background \cite{CHRISTENSEN1980480}, a coupling term between $h$ and $w$, the terms quadratic in $w$ (including a mass term and kinetic terms), and the Lagrangian multiplier terms. Similar actions were considered before in \cite{Borissova:2023yxs,Borissova:2025frj} (however in this paper we do not split the purely area metric fluctuation $w_{\ab\mn}$ into self- and anti-selfdual parts, and we do not consider terms in the action that include the Levi-Civita Tensor). Also, the term proportional to $K_2$, which is a type of kinetic term but with the indices contracted in an unusual way, was not considered in earlier papers to our knowledge, but we include it anyway out of interest. Further, we ignore any possible boundary terms.
\\

Taking the variation of \eqref{Lfull} with respect to $h^\mn$ reproduces \eqref{var_h_eq} from the main text. Taking the variation with respect to $w^{\alpha\mu\beta\nu}$ will give a second, rank-4 tensorial equation of motion which we now derive. Firstly, from the variation of the terms in $\mL_{w^2}[w]$ we obtain the "naive" expression (as defined in the preceding section)
\begin{align}
    X_{\alpha\mu\beta\nu}=
    -2m^2 w_{\alpha\mu\beta\nu}
    +2K_1 \nabla_\mu\nabla^\mu w_{\alpha\mu\beta\nu}
    +\frac{K_2}{4}\IS{\nabla_\nu\nabla_\delta w_{\alpha\mu\beta}^{\ \ \ \ \delta}}{\alpha\mu}{\beta\nu}.
\end{align}
where the notation defined in \eqref{notat} was used again. It is interesting to note that for this specific expression, $\epsilon^{\gamma\delta\rho\sigma}X_{\gamma\delta\rho\sigma}=0=X^{\gamma\delta}_{\ \ \gamma\delta}$ and only the term proportional to $K_2$ is not automatically traceless, hence the terms arising from integrating out the Lagrange multipliers (as was done in \eqref{physical_EOMs}) are going to be relatively simple. Taking the variation of the term $\mL_{h\text{-}w}$ with respect to $w^{\alpha\mu\beta\nu}$ gives the expression 
\begin{align}
    X_{\alpha\mu\beta\nu}=-\frac{1}{16}\IS{\nabla_\alpha\nabla_\beta h_{\mu\nu}}{\alpha\mu}{\beta\nu}.
    \label{ddh}
\end{align}
Upon integrating out the Lagrange-multipliers, the physical equation of motion will only depend on the traceless cyclic part of this expression (see again the derivation of \eqref{physical_EOMs}). This expression is rather ugly when derived and written down in a brute-force manner, however we can make the following observations: The resulting expression should 
\begin{itemize}
\item be linear in $h_\mn$ and depend on second order derivatives of it,
\item respect the gauge symmetry, i.e.~vanish for $h_\mn=\nabla_\mu\xi_\nu+\nabla_\nu\xi_\mu$ (notably, the expression in \eqref{ddh} by itself is not gauge invariant on an AdS background, where covariant derivatives do not commute in general),
\item be a rank-4 tensor with the same index-symmetries as an area metric,
\item and be traceless and cyclic. 
\end{itemize}
These conditions together suggest that the resulting expression should be proportional to $\delta^{(h)}_1 C_{\alpha\mu\beta\nu}$, the first order change of the bulk Weyl-tensor due to the metric fluctuation $h_\mn$, and only the prefactor remains to be determined. This also implies that, ignoring boundary terms, the coupling term between $h$ and $w$ can be rewritten as 
\begin{align}
    \mL_{h\text{-}w}[h,w]&\propto \left(\delta^{(h)}_1 C^{\alpha\mu\beta\nu}\right)w_{\alpha\mu\beta\nu}.
\end{align}
Determining the correct proportionality factor requires the appropriate sorting of covariant derivatives using the Ricci identity to commute them, and the explicit expression of the Riemann tensor of the background AdS metric in terms of the metric tensor.   
\\

All in all, after integrating out the Lagrange multipliers, the two physical equations of motion read 
\begin{align}
    \frac{\delta \mS}{\delta h^\mn} = 0 \Rightarrow 0&= 
    {E_\mn}
   {\ \ -\ \ \frac{1}{2}\nabla^{\rho}\nabla^{\lambda} w_{\mu\rho\nu\lambda}},
   \label{w_eq_1}
    \\
       \frac{\delta \mS}{\delta w^{\alpha\mu\beta\nu}} = 0 \Rightarrow 0&= 
      {
      \frac{1}{4}\delta^{(h)}_1 C_{\alpha\mu\beta\nu}
      }
     -2m^2 w_{\alpha\mu\beta\nu}
    +2K_1 \nabla_\rho\nabla^\rho w_{\alpha\mu\beta\nu}
    +\frac{K_2}{4}\IS{\nabla_\nu\nabla_\delta w_{\alpha\mu\beta}^{\ \ \ \ \delta}}{\alpha\mu}{\beta\nu} 
    -\frac{K_2}{4}\IS{g_{\mu\nu}\nabla^\gamma\nabla^\delta w_{\alpha\gamma\beta\delta}}{\alpha\mu}{\beta\nu} .
     \label{w_eq_2}
\end{align}

\subsection{Solutions}

We leave a thorough investigation of the solutions of \eqref{w_eq_1} and \eqref{w_eq_2} for the future. However, as indicated in the main text, carrying out the holographic renormalisation procedure and demanding consistency of the holographic dictionary derived from it with the result \eqref{hwDictionary} derived from the first law of entanglement may allow us to constrain the arbitrary parameters $\{m^2,K_1,K_2\}$ in \eqref{Lfull_w2}. The very first step of this procedure requires us to identify the normalisable and non-normalisable modes of the fields of interest. 
For $h_\mn$ these are well understood \cite{deHaro:2000vlm,Skenderis:2002wp}, and for $w_{\ab\mn}$ our results \eqref{w_as_sum} and \eqref{wDictionary} indicate that the mode encoding the dual vacuum expectation value (vev $\leftrightarrow$ normalisable mode) should behave asymptotically as $w\sim1/z$. A simple scaling argument then reveals that a corresponding non-normalisable mode should behave as $w\sim1/z^4$.
Alternatively, we could deduce this from the fact that the background (AdS) metric $g_\mn$ goes as $\sim z^{-2}$ near the boundary (not coincidentially like the non-normalisable mode for $h$), while the background area metric \eqref{InducedAreametric} approaches the boundary like $\sim1/z^4$. 
We summarise this in the following table: 
\renewcommand{\arraystretch}{1.5}
   \begin{center}
    \begin{tabular}{c|c|c}
       \ \          & normalisable mode  & non-normalisable mode  \\
       \hline
       $h_\mn$      & $\sim z$\ \ \ \  & $\sim z^{-2}$  \\
       \hline
       $w_{\mn\ab}$  & $\sim z^{-1}$ & $\sim z^{-4}$  \\
       \end{tabular}
    \end{center}

Consistency between holographic renormalisation and first law of entanglement then requires first and foremost that the equations \eqref{w_eq_1} and \eqref{w_eq_2} even allow for solutions corresponding to (some of the) types indicated in the table above. 
\\

Unfortunately, the equations \eqref{w_eq_1} and \eqref{w_eq_2} are a system of coupled equations for $h_\mn$ and $w_{\ab\mn}$, and so their general solutions are hard to discuss, even under the simplifying assumptions that both $h_\mn$ and $w_{\ab\mn}$ are functions of $z$ only, and that $h_\mn$ is given in Fefferman-Graham gauge $h_{\mu z}=0$. 
We thus make one more simplifying assumption: Instead of searching for the generic solutions of the system, we only look at the subset of solutions for which the $h$- and $w-$dependent terms in \eqref{w_eq_1} and \eqref{w_eq_2} vanish individually instead of merely cancelling. I.e.~we look at the system of equations 
\begin{align}
0&= E_\mn
\label{E}
\\
0&= \delta^{(h)}_1 C_{\alpha\mu\beta\nu}
\label{C}
\\
0&=\nabla^{\rho}\nabla^{\lambda} w_{\mu\rho\nu\lambda}
\label{model1}
\\
0&=     -2m^2 w_{\alpha\mu\beta\nu}
    +2K_1 \nabla_\rho\nabla^\rho w_{\alpha\mu\beta\nu}
    +\frac{K_2}{4}\IS{\nabla_\nu\nabla_\delta w_{\alpha\mu\beta}^{\ \ \ \ \delta}}{\alpha\mu}{\beta\nu} 
    -\frac{K_2}{4}\IS{g_{\mu\nu}\nabla^\gamma\nabla^\delta w_{\alpha\gamma\beta\delta}}{\alpha\mu}{\beta\nu} .
    \label{model7}
\end{align}
Any solution to this system is trivially also a solution to \eqref{w_eq_1} and \eqref{w_eq_2}. 
Our finding is that equations \eqref{E} and \eqref{C} together only allow for solutions of the form $h\sim z^{-2}$. Equation \eqref{model1} constrains \textit{some} components of $w_{\ab\gd}$ to a from like e.g.~$w_{0101}=\frac{a_{0101}}{z^2}+\frac{b_{0101}}{z} $ while some others remain unconstrained. In contrast, \eqref{model7} only allows for solutions of the form 
\begin{align}
    w_{\ab\gd}=a_{\ab\gd}\ z^{-\frac{5}{2}-\frac{1}{2}\sqrt{\frac{132K_1+9K_2+16L^2m^2}{4K_1+K_2}}} + b_{\ab\gd}\ z^{-\frac{5}{2}+\frac{1}{2}\sqrt{\frac{132K_1+9K_2+16L^2m^2}{4K_1+K_2}}}.
\end{align}
This means that for the specific value
\begin{align}
     m^2=-\frac{6}{L^2}K_1,
\end{align}
we obtain solutions to \eqref{E} -- \eqref{model7} (and thus also to \eqref{w_eq_1} -- \eqref{w_eq_2}) of the form $h\sim z^{-2}$, $w\sim z^{-1}$, i.e.~a non-normalisable mode in $h$ and a normalisable mode in $w$ coexisting. A lot more work needs to be put into this analysis to allow for definitive statements, but it serves as an initial proof of our idea that the key to understanding a putative bulk Lagrangian for area metric gravity in AdS/CFT lies in working out the holographic renormalisation.

\section{Lanczos potentials}

\subsection{Definition}

As stated in the main text, we can follow \cite{Bampi1983} to write the purely area metric fluctuation $w_{\alpha\mu\beta\nu}$ in terms of a \textit{Lanczos potential}
\begin{align}
\ell_{\ab\gamma}=-\ell_{\ba\gamma},\  \ell_{[\ab\gamma]}=0,
\end{align}
as 
\begin{align}
w_{\ab\gd}=&\IS{\nabla_\delta \ell_{\ab\gamma}}{\ab}{\gd}+\frac{1}{3}\nabla_\mu \ell^{\nu\mu}_{\ \ \nu}\IS{g_{\alpha\gamma}g_{\beta\delta}}{\ab}{\gd}
-\frac{1}{2}\IS{g_{\alpha\gamma}\left(\nabla_\mu \ell_{\beta\ \delta}^{\ \mu}+\nabla_\mu \ell_{\delta\ \beta}^{\ \mu}-\nabla_\delta \ell_{\beta\ \mu}^{\ \mu}-\nabla_\beta \ell_{\delta\ \mu}^{\ \mu}\right)}{\ab}{\gd}
\label{wbyT}
\end{align}
Note that some factors in this equation are different from the paper \cite{Bampi1983}, but we think they are correct this way. This can be verified by checking the tracelessness of $w_{\ab\gd}$ with this ansatz.

\subsection{Lagrange multipliers}
\label{LM_for_T}

As shown in \cite{Bampi1983}, the cyclicity condition of the Lanczos potential, 
\begin{align}
    \ell_{[\ab\gamma]}=0 \Leftrightarrow \epsilon^{\ab\gamma\delta}\ell_{\ab\gamma}=0,
\end{align}
is related to the cyclicity condition \eqref{TraceW} of $w_{\ab\gd}$. Just as in section \ref{LM_for_w}, we will make sure to enforce this condition by an explicit Lagrange multiplier term in the Lagrangian: 
\begin{align}
  \mL'=\mL+\epsilon^{\ab\gamma\delta}\ell_{\ab\gamma}\lambda_\delta.
  \label{mLprime_T}
\end{align}
So if the "naive equations of motion" (derived from $\mL$ by variation w.r.t. $\ell_{\ab\gamma}$) are $Y_{\ab\gamma}$, we get from \eqref{mLprime_T} the physical equations of motion
\begin{align}
&0\equiv Y_{\ab\gamma}+\epsilon_{\ab\gamma\delta}\lambda^\delta
\\
\Rightarrow\ &0\equiv Y_{\ab\gamma}\epsilon^{\ab\gamma \rho}-6\lambda^\rho
\\
\Rightarrow\ &0\equiv Y_{\mu\nu\sigma}+\frac{1}{6}\epsilon_{\mu\nu\sigma\delta}\epsilon^{\ab\gamma \delta}Y_{\ab\gamma}
 = \frac{2}{3}Y_{\mu\nu\sigma}+\frac{1}{3}Y_{\mu\nu\sigma}-\frac{1}{3}Y_{\mu\nu\sigma}.
\end{align}

\subsection{Equations of motion}

When treating the Lanczos potential as the fundamental variable, we consider actions like 
\begin{align}
    \mL=&\mL_{EH}[h]+\mL_{h\text{-}w}[h,\ell]+\mL_{\ell^2}[\ell]+\mL_{LM}[\ell,\lambda]
    \label{Lfull_T}
\end{align}
with 
\begin{align}
\mL_{EH}[h]&=
-\frac{1}{4}\tilde{h}^\mn\Box h_{\mn}
-\frac{1}{2}\nabla^\rho\tilde{h}_{\rho\mu}\nabla^\sigma\tilde{h}_\sigma^\mu
+\frac{\Lambda}{6}\tilde{h}^\mn h_\mn
+\frac{\Lambda}{6}h^2
 \, \left(\text{with}\,\, \tilde{h}_\mn=h_\mn-\frac{1}{2}g_{\mn}h\right),
\\
\mL_{h\text{-}w}[h,\ell]&=-\frac{1}{2}h^{\ab}\nabla^{\mu} \nabla^{\nu} w_{\alpha \mu \beta \nu},
\\
\mL_{\ell^2}[\ell]&=-K_3 w_{\alpha \mu \beta \nu}w^{\alpha \mu \beta \nu}
\\
\mL_{LM}[\ell,\lambda]&= \epsilon^{\ab\gamma\delta}\ell_{\ab\gamma}\lambda_\delta.
\end{align}
Here, wherever $w_{\ab\mn}$ appears, it is understood to be expressed in terms of $\ell_{\ab\gamma}$ via \eqref{wbyT}. Consequently, a term like $w_{\alpha \mu \beta \nu}w^{\alpha \mu \beta \nu}$ which used to be a mass term in section \ref{sec::Lagrangian_EOM_w} is now a kinetic term for $\ell_{\ab\gamma}$, and we do not consider terms like $(\nabla_\delta w_{\alpha \mu \beta \nu})(\nabla^\delta w^{\alpha \mu \beta \nu})$ as they would correspond to fourth-order derivative terms for $\ell_{\ab\gamma}$. We could have added a mass-term $m^2 \ell_{\ab\gamma}\ell^{\ab\gamma}$ for the Lanczos potential, but we do not do this here for the sake of simplicity. 
\\

Taking the variation of \eqref{Lfull_T} with respect to $h^\mn$ still reproduces \eqref{var_h_eq} from the main text. Taking the variation with respect to $\ell^{\ab\gamma}$ will give a second, rank-3 tensorial equation of motion which we now derive. Firstly, from the variation of $\mL_{\ell^2}[\ell]$ we obtain the "naive" expression (as defined in the preceding section)
\begin{align}
    Y_{\alpha\beta\gamma}=16 K_3 \nabla^\delta w_{\alpha\beta\gamma\delta}.
\end{align}
Due to the cyclicity condition of $w_{\ab\mn}$, we have $\epsilon^{\ab\gamma \delta}Y_{\ab\gamma}=0$ and thus no difference between naive and physical equations arising from this term. Taking the variation of the term $\mL_{h\text{-}w}$ with respect to $\ell^{\ab\gamma}$ gives an expression which is too cumbersome to state here, and which following the main text we will explicitly refer to as $F_{\alpha\beta\gamma}$ instead of $Y_{\alpha\beta\gamma}$ from now on. 
The following observations can be stated: $F_{\alpha\beta\gamma}$
\begin{itemize}
\item is linear in $h_\mn$ and depends on third order derivatives of it,
\item is a rank-3 tensor with the same index-symmetries as a Lanczos potential,
\item is traceless ($g^{\alpha \gamma}F_{\ab\gamma}=0$) 
\item and cyclic ($\epsilon^{\ab\gamma \delta}F_{\ab\gamma}=0$). 
\end{itemize}
The last fact means that for this term again, the Lagrange multiplier wasn't explicitly needed and no difference between "naive" and physical equations of motions arises.  
These conditions are not stringent enough to suggest a unique answer, e.g.~$\nabla^\nu\delta^{(h)}_1 C_{\alpha\beta\gamma\nu}$ would have been a possible candidate. Instead, as indicated in the main text, a lengthy calculation reveals that this can be written in terms of derivatives of the Einstein equations: 
\begin{align}
    F_{\ab\gamma}\equiv
    \frac{1}{3}g_{\alpha\gamma}\nabla_\beta E^\mu_\mu
    -\frac{1}{3}g_{\beta\gamma}\nabla_\alpha E^\mu_\mu
    -\nabla_\beta  E_{\alpha\gamma}
    +\nabla_\alpha E_{\beta\gamma}
    ,
\end{align}
Proving this requires again the appropriate sorting of covariant derivatives using the Ricci identity to commute them, and the explicit expression of the Riemann tensor of the background AdS metric in terms of the metric tensor.  
All in all, the equations of motion resulting from \eqref{Lfull_T} read 
\begin{align}
    \frac{\delta \mS}{\delta h^\mn} = 0 \Rightarrow 0&= 
    {E_\mn}
   {\ \ -\ \ \frac{1}{2}\nabla^{\rho}\nabla^{\lambda} w_{\mu\rho\nu\lambda}},
   \label{Lanczos_rank2}
    \\
       \frac{\delta \mS}{\delta \ell^{\ab\gamma}} = 0 \Rightarrow 0&= 
    \frac{1}{3}g_{\alpha\gamma}\nabla_\beta E^\mu_\mu
    -\frac{1}{3}g_{\beta\gamma}\nabla_\alpha E^\mu_\mu
    -\nabla_\beta  E_{\alpha\gamma}
    +\nabla_\alpha E_{\beta\gamma}
    +16 K_3 \nabla^\delta w_{\alpha\beta\gamma\delta} .
    \label{Lanczos_rank4}
\end{align}
These equations are interesting, because in the "Einsteinian" limit $w\rightarrow 0$ (where the geometry reduces to the ordinary length-metric geometry of Einsteinian gravity), we are left with 
\begin{align}
 0&= 
    {E_\mn},
    \\
 0&= 
    \frac{1}{3}g_{\alpha\gamma}\nabla_\beta E^\mu_\mu
    -\frac{1}{3}g_{\beta\gamma}\nabla_\alpha E^\mu_\mu
    -\nabla_\beta  E_{\alpha\gamma}
    +\nabla_\alpha E_{\beta\gamma},
\end{align}
and thus, the second equation being redundant, we re-obtain not only Einsteinian geometry, but (linearised) Einsteinian gravity in this limit. This was not the case when treating $w$ as the fundamental variable in section \ref{sec::Lagrangian_EOM_w}. 

\subsection{Solutions}

Again, we very briefly discuss some solutions of the equations \eqref{Lanczos_rank2} and \eqref{Lanczos_rank4}. First of all, because $w$ is given by covariant derivatives of $\ell$ in \eqref{wbyT}, we might naively expect $w\sim z^{-1}\Leftrightarrow \ell\sim const.$ and 
$w\sim z^{-4}\Leftrightarrow \ell\sim z^{-3}$. The former condition works because the Christoffel-symbols will introduce factors proportional to $z^{-1}$, so an ansatz like $\ell\sim \log(z)$ would not work. Thus, we might very naively expect to find solutions of the form:
\renewcommand{\arraystretch}{1.5}
   \begin{center}
    \begin{tabular}{c|c|c}
       \ \          & normalisable mode  & non-normalisable mode  \\
       \hline
       $h_\mn$      & $\sim z$ & $\sim z^{-2}$  \\
       \hline
       $\ell_{\ab\gamma}$  & $\sim const. $ & $\sim z^{-3}$  \\
       \end{tabular}
    \end{center}

Making the same assumptions as previously (only $z$-dependence and Fefferman-Graham gauge for $h_\mn$), we also once again replace the coupled equations \eqref{Lanczos_rank2} and \eqref{Lanczos_rank4} with the more restrictive system of non-coupled equations where all relevant terms vanish individually, i.e.
\begin{align}
 0&=  E_\mn\ \ \  \ \ \ \ \ \ \left(\Rightarrow 
 \frac{1}{3}g_{\alpha\gamma}\nabla_\beta E^\mu_\mu
    -\frac{1}{3}g_{\beta\gamma}\nabla_\alpha E^\mu_\mu
    -\nabla_\beta  E_{\alpha\gamma}
    +\nabla_\alpha E_{\beta\gamma}=0 \ \ \text{  automatically}\right),
 \\
 0&= \nabla^\delta w_{\alpha\beta\gamma\delta}\ \ \  \left(\Rightarrow \nabla^{\rho}\nabla^{\lambda} w_{\mu\rho\nu\lambda}=0 \ \ \text{  automatically}\right).
 \label{1st_order_w}
\end{align}
Solutions of these equations will at least form a subset of the solutions of \eqref{Lanczos_rank2} and \eqref{Lanczos_rank4}. For $h_\mn$, this subset includes \textit{all} solutions to Einstein's equations (including both normalisable and non-normalisable modes). For $w$ respectively $\ell$, we find that \eqref{1st_order_w} allows for solutions of the form $w\neq 0;\ w\sim z^{-1} \Leftrightarrow \ell\sim const.$ (the expected normalisable mode). There are no possible solutions with $w\sim z^{-4}$. However, there is also the solution $w_{\ab\mn}=0$ which, thanks to the somewhat involved definition \eqref{wbyT}, can be obtained from non-trivial solutions of the form $\ell\sim z^{-3}$ (in addition to a large remaining degree of degeneracy, discussed in \cite{Takeno,Bampi1983,Vishwakarma:2020yvo,Gopal:2021lax}). It thus appears that all the modes conjectured in the above table to be important for holographic renormalisation of this model can indeed exist. The model \eqref{Lfull_T} with $\ell_{\ab\gamma}$ as fundamental variable certainly will warrant further study in the future.

\end{widetext}
\bibliography{references}

\end{document}